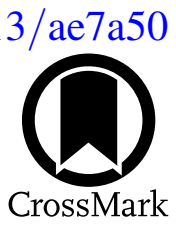

# TESS's First Bound Microlensing Planet—A Binary Microlensing Event Revealing a Planetary Companion toward the Galactic Plane

Mallory Harris[1], Diana Dragomir[1], Etienne Bachelet[2], Michael Fausnaugh[3], and Samson Johnson[4]
[1] Department of Physics and Astronomy, University of New Mexico, Albuquerque, NM 87106, USA
[2] Université Marie et Louis Pasteur, CNRS, Institut UTINAM UMR 6213, Besançon, France
[3] Department of Physics and Astronomy, Box 41051, Texas Tech University, Lubbock, TX 79409, USA
[4] Department of Astronomy, The Ohio State University, 140 West 18th Avenue, Columbus, OH 43210, USA

## Abstract

We report the discovery of Gaia23bra b, the first gravitationally bound microlensing planet detected by the Transiting Exoplanet Survey Satellite (TESS). Initially flagged as a single-lens event by the Gaia Science Alerts system, Gaia23bra was serendipitously observed by TESS over two consecutive sectors. During those TESS sectors, the light curve of the event displayed caustic-crossing features characteristic of a binary-lens event. The joint modeling of Gaia and TESS photometry with `pyLIMA`, supplemented by stellar parameter inference using `pyLIMASS`, suggests a K dwarf ($M_L = 0.79^{+0.19}_{-0.17}\, M_\odot$) hosting a Jovian planet with $M_P = 1.63^{+0.42}_{-0.38}\, M_{\rm Jup}$ at a projected separation of $a_{\perp,\rm min} \approx 4.8$ au. This result underscores the synergy between high-cadence photometry and long-baseline monitoring for robust microlensing characterization. Its location along the Galactic Plane highlights TESS's unexpected capacity for microlensing science through its all-sky coverage and its potential to detect planets in regions beyond the Galactic Bulge.

## 1. Introduction

The 2010 and 2020 Astronomy and Astrophysics Decadal Surveys identified a persistent gap in our census of exoplanet systems at orbital distances comparable to and beyond Earth's orbit (National Research Council 2010; National Academies of Sciences, Engineering, & Medicine 2021). This gap arises due to the fact that most known exoplanets have been discovered with the transit or radial velocity methods, both of which are less sensitive to longer-period planets. In contrast, the microlensing method of planet detection—where the gravitational field of a foreground star acts as a lens, magnifying the light of a background star when the systems align near the observer's line of sight, with planets producing characteristic perturbations to this signal (B. Paczynski 1986a)—is sensitive to planets at wider separations, including beyond the ice line and even free-floating planets (D. P. Bennett & S. H. Rhie 2002). Microlensing thus provides a powerful tool for probing planetary populations that remain inaccessible to transit and radial velocity surveys (D. Suzuki et al. 2016; R. Poleski et al. 2021; W. Zang et al. 2025).

The upcoming Nancy Grace Roman Space Telescope (Roman; D. Spergel et al. 2015), slated for launch in 2026, will be the first mission designed with microlensing as a primary goal, and is expected to discover hundreds of exoplanets through a dedicated survey of the Galactic Bulge (M. T. Penny et al. 2019; S. A. Johnson et al. 2020). While Roman represents a major advance, previous space-based observatories have already demonstrated the viability of microlensing science using space-based photometric observations. For example, Kepler's K2 Campaign 9 conducted the first space-based microlensing survey of the Bulge (A. Gould & K. Horne 2013; S. B. Howell et al. 2014; C. B. Henderson et al. 2016), detecting several events, including the first space-discovered microlensing planet, K2-2016BLG-0005Lb (D. Specht et al. 2023). Similarly, the Spitzer Space Telescope contributed by measuring microlensing parallax and constraining lens masses (S. Dong et al. 2007; A. Udalski et al. 2015), establishing a precedent for leveraging general-purpose photometric missions for microlensing science (C. A. Beichman & D. Deming 2018).

The primary mission of the Transiting Exoplanet Survey Satellite (TESS), launched in 2018 April, is to find transiting planets around nearby bright stars (G. R. Ricker et al. 2015), but its high-cadence, wide-field observations have enabled applications well beyond its original scope (J. N. Winn 2024). Prior to this work, M. Kunimoto et al. (2025) reported a candidate microlensing event in TESS data, though subsequent analyses questioned its interpretation as such (P. Mróz 2023; H. Yang et al. 2024). Regardless of its classification, that study highlighted the possibility of microlensing detections with TESS—a capability we confirm here, through the first definitive discovery of a gravitationally bound microlensing planet, Gaia23bra b. Although TESS was not designed for microlensing, and faces challenges such as short observing windows and large pixel scale, this detection demonstrates how the combination of high-cadence photometry with complementary long-baseline observations from Gaia or ground-based facilities can enable new microlensing detections and characterization. Further, with TESS's observing strategy sampling the entire sky over the course of its continuing mission, it has the potential to provide observations of microlensing events outside the Galactic Bulge.

**Table 1**
Properties of the Star Gaia DR3 5252141822116891648 from Gaia DR3 and the Gaia Science Alert

| Parameter | Value |
|---|---|
| R.A. (deg) | 155.62945 |
| Decl. (deg) | −63.26392 |
| $l$ (deg) | 287.07379 |
| $b$ (deg) | −5.0738 |
| $\mu_{\mathrm{RA}}$ (mas yr$^{-1}$) | $-3.2543 \pm 0.2610$ |
| $\mu_{\mathrm{Dec}}$ (mas yr$^{-1}$) | $1.5670 \pm 0.2564$ |
| RUWE | 0.984 |
| Parallax, $\pi$ (mas) | $-0.0143 \pm 0.2095$ |
| $G_{\mathrm{mag}}$ | $19.3824 \pm 0.0027$ |
| $G_{\mathrm{BP,mag}}$ | $19.8231 \pm 0.0478$ |
| $G_{\mathrm{RP,mag}}$ | $18.5553 \pm 0.0367$ |
| $G_{\mathrm{BP,mag}} - G_{\mathrm{RP,mag}}$ | $1.2678 \pm 0.0603$ |
| Alerted Magnitude ($G_{\mathrm{mag}}$) | 18.64 |
| Historical Magnitude ($G_{\mathrm{mag}}$) | 19.44 |
| Historical StdDev ($G_{\mathrm{mag}}$) | 0.04 |

**Note.** The event triggered an alert when it brightened by 0.8 mag. The Renormalized Unit Weight Error (RUWE) for this star is close to 1, indicating that the source is likely a single star, rather than a multistar system.

Recent studies have shown that microlensing events occur along the Galactic Plane (P. Mróz et al. 2020; Ł. Wyrzykowski et al. 2023), suggesting that wide-area surveys can complement Bulge-focused programs. However, the feasibility of systematic microlensing searches with TESS remains limited without additional observations from another instrument over a longer timescale, such as Gaia. Our result showcases a unique synergy between high-cadence photometric missions and long-baseline monitoring, and presents a strategy that can in the future be employed by TESS and Roman in conjunction with the Vera Rubin Survey.

In this Letter, we place these results into context through a description of the observations, modeling framework, and physical interpretation of the Gaia23bra microlensing event. Section 2 summarizes the Gaia and TESS datasets. Section 3 presents the microlensing models used for the Gaia-only, TESS-only, and joint fits. Section 4 describes the inferred properties of the source, lens, and planetary companion, and Section 5 discusses the broader implications for microlensing searches with TESS and future missions.

## 2. Observations

### 2.1. Gaia Observations

The Gaia satellite has provided high-precision astrometry and photometry since its launch in 2013, delivering repeated measurements of the sky through the end of its mission in early 2025 (Gaia Collaboration et al. 2016). The Gaia Science Alerts system was developed to identify transient brightening in near-real-time and to enable timely follow-up observations (S. T. Hodgkin et al. 2021).

The microlensing event Gaia23bra was first identified through Gaia Science Alerts on 27 April 2023, with the official alert published on 5 May 2023. The event occurred near the Galactic Plane, at galactic coordinates $l = 287^{\circ}.07379$, $b = -5^{\circ}.0738$ (see Table 1). The alert included a preliminary light curve spanning from 2014 November 9 to 2023 December 24, with magnitudes in the Gaia G band. These magnitudes were derived from an initial calibration of Gaia photometry, and each observation had an effective exposure time of 4.42 s (C. Crowley et al. 2016). We applied the error model described in K. Kruszyńska et al. (2022; see the equation between their Equations (2) and (3)), which accounts for instrumental noise and calibration effects in Gaia time-series photometry.

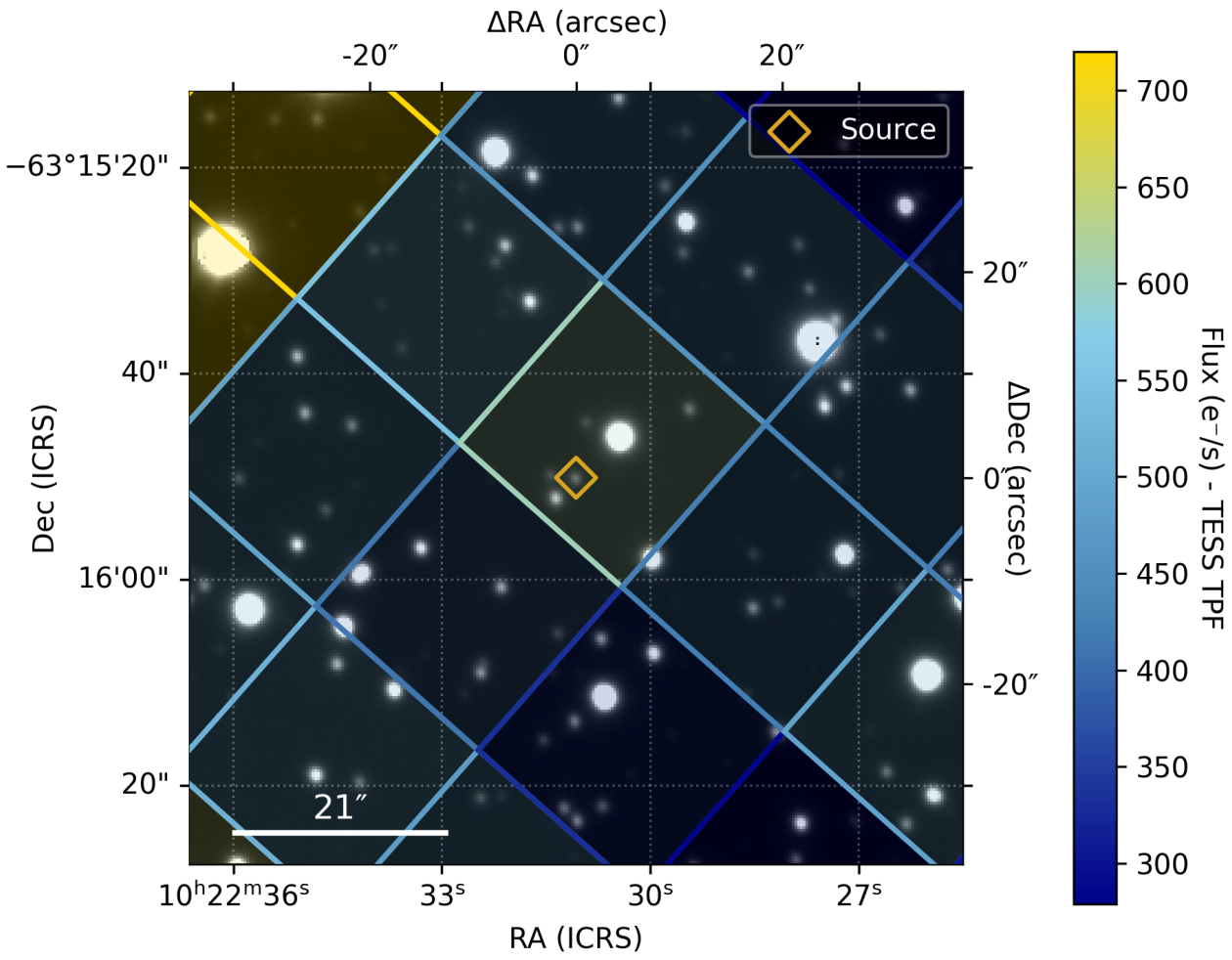

**Figure 1.** Overlay of the TESS pixel grid on a DECaPS2 cutout centered on the source star of the Gaia23bra event, marked by the empty gold diamond. The diamonds trace the 21" x 21" TESS pixels, with their colors indicating the brightness in the TESS pixels. The plot shows the source's location within a single TESS pixel and the surrounding crowding.

Prior to the microlensing event, the source star maintained a stable $G$-band baseline magnitude of 19.44 ($\sigma = 0.04$). A gradual rise in flux became apparent in 2023 mid-March, culminating in a peak on 2023 April 27 at magnitude 18.64—a $20\sigma$ deviation from the baseline. The flux declined toward baseline levels by 2023 June 30, with Gaia obtaining five measurements during this interval. After 2023 June 30, the light curve shows only a shallow upward trend through the final data point on 2023 December 24, remaining within $2\sigma$ of the baseline.

### 2.2. TESS Observations

Our collaboration mines all Gaia transient events that overlap with simultaneous TESS sectors, including microlensing events. The timing of Gaia23bra coincided with TESS Extended Mission Year 5, which included partial coverage of both hemispheres. This coverage spanned Sectors 56–60 in the north and Sectors 61–69 in the south. During this period, full-frame images were obtained at a cadence of 200 s.

TESS observed Gaia23bra during Sectors 63 and 64, from 10 March 2023 to 4 May 2023. The TESS light curve includes the increase in flux observed by Gaia. Unfortunately, no clear baseline was observed in by TESS, as the event was already underway at the start of Sector 63.

Due to TESS's relatively coarse spatial resolution of 21" per pixel, compared to Gaia's $0''.059$ per pixel in the scanning direction (illustrated in Figure 1), we extracted the TESS light curve using difference imaging (C. Alard & R. H. Lupton 1998; C. Alard 2000). Our method is the same as the image-subtraction technique described in Appendix A of M. M. Fausnaugh et al. (2023c). First, we built a reference image using a median stack of 20 images with low backgrounds from the TESS sector. We then solved for a smoothing kernel that matches the point-spread function (PSF) of a given image to the reference image, before subtracting the two images, leaving only the variable flux in each pixel. For TESS, this procedure corrects for the main systematic errors caused by image shifts, pointing jitter, and thermal effects.

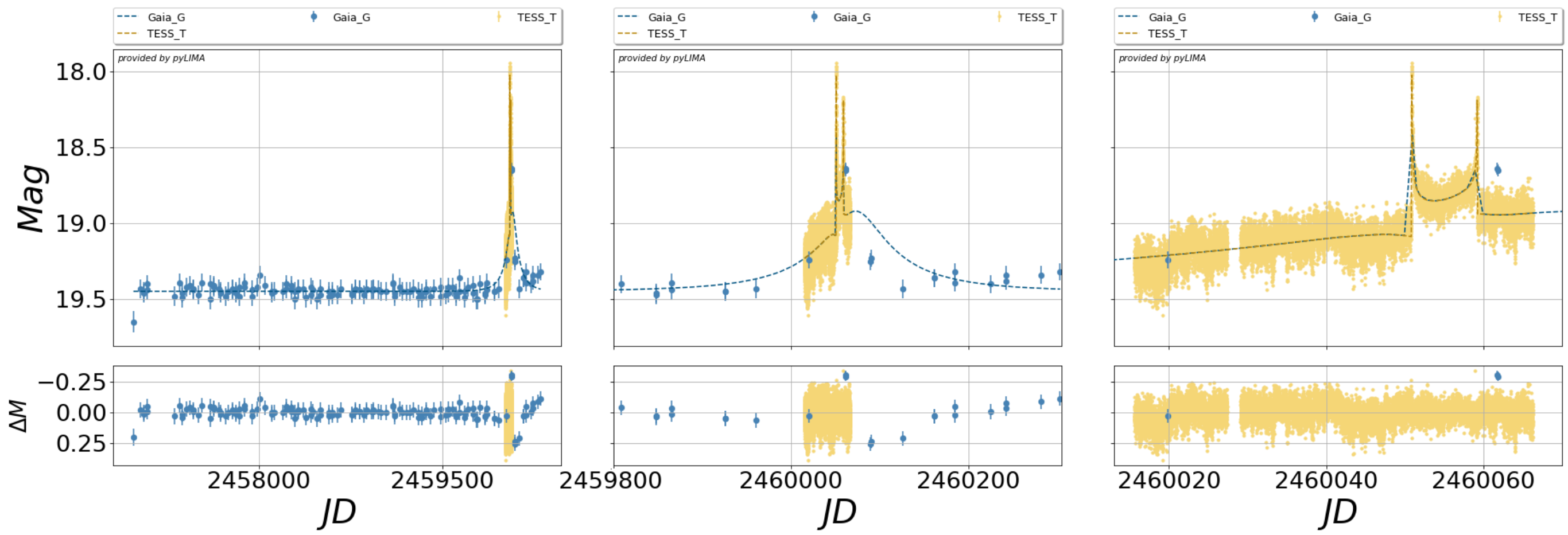

**Figure 2.** Photometric observations of the Gaia23bra microlensing event from Gaia (blue) and TESS (gold), with best-fit `pyLIMA` models overlaid as dashed lines in dark blue and dark gold, respectively. The panels from left to right show: Gaia's full baseline coverage of the event; the modeled light curve illustrating the event timescale; and TESS observations capturing high-cadence, caustic-crossing features.

After acquiring the difference images, we used the models of the TESS PSF[5] smoothed with the optimal kernels from the image subtraction to perform forced photometry at the location of Gaia23bra. We also corrected residual backgrounds by subtracting the median pixel value in an annulus (inner/outer radii of 4 and 8 TESS pixels, respectively) from the integrated PSF flux from each difference image. Additional details of the difference-imaging procedure and forced photometry are given in M. M. Fausnaugh et al. (2023c).

We used separate reference images for Sector 63 and Sector 64. The reference image sets the zero-point of the differential light curve, so it is necessary to align the two light curves to the same flux scale. We "stitched" the sectors together using a smooth Gaussian Process model. We first binned the light curves to 1.5 days, which decreases the effect of the measurement noise on the Gaussian Process model and speeds up the fitting procedure. We used a squared-exponential kernel for the Gaussian Process model, ensuring that the underlying variations were smooth and preserved the general shape of the microlensing event. We then searched a grid of shifts for Sector 64, optimizing the correlation length with respect to the data for each shift. We selected the shift that produces the lowest $\chi^2$ value of the data residuals relative to the Gaussian Process interpolation. The absolute flux calibration is set by shifting Sector 63 so that the average flux matches the Gaia$G_{RP}$ magnitude of 18.56.

Figure 2 depicts the TESS light curve of the Gaia23bra event in gold. The light curve gradually increases in brightness, before displaying peaks characteristic of microlensing caustic crossings on Julian dates 2460050.88 (2023 April 16) and 2460059.22 (2023 April 24). The curvature near the peaks is indicative of either finite-source effects or limb darkening from the source star. After the caustic crossing, the observed brightness continues to rise through the end of Sector 64 on 2023 May 4.

## 3. Analysis

Maximizing the scientific return from an event requires identifying and fitting the most appropriate microlensing model. For the Gaia23bra event, we applied the `pyLIMA` software to determine the best-fit model (E. Bachelet et al. 2017). This section begins with a mathematical overview of how microlensing observables relate to physical parameters, followed by a description of the `pyLIMA` analysis we conducted using the Gaia and TESS observations.

### 3.1. Microlensing Geometry

Microlensing events occur when the gravity of a foreground lens object magnifies the light of a background source star as the two nearly align along the observer's line of sight (B. Paczynski 1986a). The observed flux at time $t$ can be described by

$$F(t) = F_{\rm S} A(u) + F_B, \tag{1}$$

where $F_{\rm S}$ is the flux of the source star, $F_B$ is the blend flux from unrelated nearby objects, and $A(u)$ is the magnification factor caused by the lens (B. Paczyński 1986b; A. Gould 1996).

For the idealized point-source, point-lens (PSPL) case, the magnification has the analytic form derived by B. Paczynski (1986a), and the resulting light curve is called a Paczyński curve. This magnification is

$$A(u) = \frac{u^2 + 2}{u\sqrt{u^2 + 4}}, \tag{2}$$

where $u$ is the angular separation between the lens and source, $\theta(t)$, normalized by the Einstein radius, $\theta_E$, (B. Paczyński 1986b).

The evolution of $u$ over time can be described in measurable quantities as

$$u(t) = \sqrt{\left(\frac{(t - t_0)}{t_{\rm E}}\right)^2 + u_0^2}, \tag{3}$$

[5] https://archive.stsci.edu/missions/tess/models/prf_fitsfiles/

where $t_0$ is the time at closest approach and $u_0$ is the minimum lens–source separation, both of which coincide with the peak magnification of the event.

For a PS model, this maximum flux formally diverges as $u_0 \rightarrow 0$ and $A(u) \rightarrow \infty$. In practice, the finite size of the source star rounds the Paczyński curve peak as well as the binary-lens caustic features, removing the unphysical divergence (R. J. Nemiroff & W. A. D. T. Wickramasinghe 1994). These effects introduce the normalized source size $\rho = \theta_*/\theta_E$, which can be measured when the light curve resolves the caustic or peak structure (C. Han & A. Gould 1997).

The Einstein crossing time in Equation (3) can be calculated as

$$t_{\rm E} = \frac{\theta_E}{\mu_{rel}}, \quad (4)$$

where $\mu_{rel}$ is the proper motion of the lens star relative to the source. The Einstein radius, $\theta_E$, may be determined geometrically through the angular source size $\theta_*$:

$$\theta_E = \frac{\theta_\star}{\rho}. \quad (5)$$

It can also be expressed in terms of the lens mass $M_{\rm L}$ and the relative parallax $\pi_{rel}$:

$$\theta_E = \sqrt{\kappa M_{\rm L} \pi_{rel}}, \quad (6)$$

wherein $\kappa = 4G/(c^2 {\rm AU}) = 8.14 {\rm mas}/M_\odot$ (A. Gould 1996).

The relative parallax $\pi_{rel}$ is a function of the lens and source distances,

$$\pi_{rel} = \pi_L - \pi_S = \frac{\rm AU}{D_L} - \frac{\rm AU}{D_S}, \quad (7)$$

and enters into the microlensing parallax parameter $\pi_E$, defined through

$$\pi_E \equiv \frac{\rm AU}{\tilde{r}_E} = \frac{\pi_{\rm rel}}{\theta_E}. \quad (8)$$

where AU denotes the Earth–Sun baseline distance, and $\tilde{r}_E$ is the Einstein radius projected onto the observer plane. The microlensing parallax is a two-dimensional vector with components $(\pi_{E,N},\ \pi_{E,E})$ in equatorial coordinates.

$\pi_E$ is essential to determining the lens mass and distance, as it removes the mass–distance degeneracy through

$$M_{\rm L} = \frac{\theta_E}{\kappa \pi_E}. \quad (9)$$

The parallax signal can be constrained through observations from a second observatory located at a large projected separation from the primary one (A. Gould 1996).

In a binary-lens scenario where the secondary object distorts the shape of the light curve, the mass ratio and separation between the two objects create caustics—regions of high magnification produced by their combined gravitational fields (R. Blandford & R. Narayan 1986; S. Mao & B. Paczynski 1991; A. Gould & A. Loeb 1992). The shapes of these caustics and the resulting light curves are solely dependent on the mass ratio of the objects in the binary ($q$) and the projected angular separation ($s$) between the two bodies normalized by the Einstein radius. Both of these quantities can be derived through fitting these light-curve distortions (A. Gould & A. Loeb 1992). The total lens mass and Einstein radii are thus crucial quantities for fully solving the system and determining the type and orbit of a microlensing planet around its host.

#### *3.1.1. Flux Treatment for Gaia and TESS*

For Gaia observations, the source is sufficiently well resolved to allow the `pyLIMA` modeling to meaningfully split the observed flux into a source component and a blend component, represented as $f_{\rm S}$ and $f_{\rm B}$, respectively. These parameters ideally correspond to the physical source flux and the summed flux of nearby unresolved stars. Gaia was able to sample the baseline outside the microlensing event, where $A(u) = 1$ and the baseline flux is

$$F_{\rm baseline} = F_{\rm S} + F_{\rm B}. \quad (10)$$

The source flux can then be expressed as a magnitude, using

$$f_{\rm S} = 10^{-0.4(G_{\rm S} - zp)}, \quad (11)$$

following E. Bachelet et al. (2017), where $zp$ is the zero-point adopted for the flux normalization. The value of $zp$ depends on the photometric system and definition of the flux units, which for Gaia corresponds to $zp_{\rm Gaia} = 25.6874 \pm 0.0028$ (M. Riello et al. 2021). In this analysis, the zero-point was arbitrary, as the input light curves are provided in Gaia G-band magnitudes and a consistent $zp$ is used for both the magnitude-to-flux and flux-to-magnitude conversions, ensuring that the derived magnitudes remain on the $G$-band scale. This step depends on the model separating $f_{\rm S}$ and $f_{\rm B}$ correctly; the Gaia resolution is sufficiently high that this decomposition is well constrained.

TESS observations require a different approach. With a 21″ pixel scale, the target is heavily blended with many neighboring stars, such that the blend flux is much larger than the source. Although $f_{\rm S}$ and $f_{\rm B}$ are included in the microlensing model, in the TESS data, their individual values are poorly constrained, as the baseline flux level is not well determined. As a result, these observations primarily serve to constrain the change in flux during the microlensing event rather than the absolute flux. We therefore adopt the pixel-lensing formulation (A. Gould 1996)

$$\Delta F(t) \approx F_{\rm S}\,[A(u) - A(u_{\rm ref})]. \quad (12)$$

If a true baseline image were available, we would have $A(u_{\rm ref}) \approx 1$. In our case, the event had already started when Sector 63 began, so $A(u_{\rm ref}) > 1$. This does not prevent the TESS data from being used, as the blend flux is constant over each sector and our Gaussian Process stitching places both sectors onto a consistent relative flux scale. As a result, the TESS light curve still provides strong constraints on the shape of the microlensing event, even though its absolute source and blend fluxes cannot be measured.

The lower sensitivity of TESS has direct consequences for the portion of the microlensing signal TESS can detect. TESS full-frame images do not strongly constrain the ingress and egress phases of microlensing events, so only the region near the peak contributes significant signal. As a result, TESS photometry alone tends to recover shorter apparent Einstein timescales than the intrinsic event duration. This behavior is expected in pixel-lensing regimes and underscores the need for the long-baseline Gaia data to anchor the full light-curve shape.

### 3.2. Gaia Fit

The Gaia observations of Gaia23bra provide baseline measurements and the resolution to separate the source and blend fluxes (Section 3.1.1). Gaia observed the smooth, long-timescale variation of the event and, due to its sparser cadence, does not resolve the caustic-crossing structure seen in the TESS data. We thus modeled the resulting data with a traditional Paczyński curve or PSPL model. This fit served to constrain the timescale of the overall event as well as the source and blend flux parameters.

For this model, we initiated a three-step fitting sequence using `pyLIMA`'s available routines. We began with a Differential Evolution (DE) algorithm to perform a global exploration of the parameter space, ensuring convergence toward the best-fit region. This was followed by a Trust Region Reflective (TRF) optimization to refine the solution locally. Finally, we applied a Markov Chain Monte Carlo (MCMC) fit, using the `emcee` package (D. Foreman-Mackey et al. 2013), to sample the posterior distributions and quantify uncertainties. We repeated the TRF $\rightarrow$ MCMC steps for every subsequent model; the DE stage was only needed for the initial Gaia-only and TESS-only fits, to explore the parameter space broadly and mitigate sensitivity to the initial conditions.

For completeness, we also ran a Gaia-only finite-source, point-lens (FSPL) fit. As expected from the sparse cadence and lack of finite-source structure, this model was unconstrained and collapsed to the PSPL solution; it is included in Table 2 for comparison only. The well-constrained PSPL parameters in Table 2 then provided the context needed for the later joint modeling.

### 3.3. TESS Fit

To fit the TESS observations, we initially used only the Sector 64 data, as Sector 63 captured neither a clear baseline nor the caustic-crossing structure. We first fit this data with a point-source, binary-lens (PSBL) model in `pyLIMA`, which serves as an initial exploration of the parameter space and captures the overall morphology of the event. After applying the same three-step fitting sequence described in Section 3.2, we found that the PSBL model captured the overall shape of the event but did not reproduce the curvature of the caustic peaks. To better match the observed morphology, we switched to a uniform-source, binary-lens (USBL) model, which reproduced the finite-source effects visible in the TESS photometry.

Upon transitioning to non-point-source, binary-lens model regimes, we binned the TESS light curve to 1 hr intervals to make the parameter-space exploration computationally feasible. As the caustic-crossing peaks each spanned roughly 5 hr, this binning preserves their overall morphology at the relevant timescale. The model parameters are inferred from the binned data, while the resulting best-fit models are applied to the unbinned light curve to compute the $\chi^2$ values presented in Table 2, ensuring that the fit quality is assessed using the full temporal resolution of the observations.

Upon achieving a well-constrained fit to the Sector 64 light curve, we then incorporated Sector 63 data to refine the event timescale ($t_{\mathrm{E}}$) and impact parameter ($u_0$). We aligned and concatenated the two sectors using a Gaussian Process model, as described in Section 2.2. The resulting fit parameters are listed in Table 2 and served as the initial guess going into the joint Gaia+TESS modeling.

### 3.4. Joint TESS + Gaia Fit

After fitting the Gaia and TESS light curves separately, we proceeded to a joint fit combining both datasets. We began imploying the USBL model derived from the TESS-only fit and used it as the initial configuration for the joint-fitting sequence.

As the TESS observations clearly demonstrated finite-source effects at the caustic crossings, it was necessary to incorporate limb darkening into the model. As such, we transitioned to a finite-source, binary-lens (FSBL) configuration. To determine the linear limb-darkening coefficients for the source star, we adopted an iterative approach. We first used the USBL solution (specifically, the $t_E$, $\rho$, and $f_{\mathrm{S}}$ values) to derive an initial set of host-star properties using `pyLIMASS`—a stellar parameter estimation tool that infers physical properties from microlensing observables (E. Bachelet et al. 2024). From the resulting $T_{\mathrm{eff}}$, $\log g$, and [Fe/H] values, we found initial linear limb-darkening coefficients. We then incorporated these coefficients into a subsequent `pyLIMA` fit to the observations and repeated this process until the stellar parameters and limb-darkening coefficients converged.

For the TESS observations, the coefficients were selected from the A. Claret (2017) tables using the inferred stellar parameters. For the Gaia observations, `pyLIMA` instead computed linear limb-darkening coefficients from the stellar parameters within its supported filter set, corresponding to a broadband approximation of the Gaia photometric response.

Once the limb-darkening coefficients converged, we adopted the FSBL model for the remainder of the joint `pyLIMA` fits.

#### 3.4.1. Investigating Degenerate Solutions

Upon employing the binary-lens model for both datasets, we observed a bimodality in the fits, indicative of two degenerate solutions. This behavior is common in microlensing analyses, where structurally similar configurations often produce comparably good fits (N. Kains et al. 2009). To investigate this, we used two complementary approaches to search for additional minima in the binary-lens parameter space. The first approach relied on a direct mapping of $\chi^2$ across a grid in ($\log s$, $\log q$) (N. Kains et al. 2009; Y. Tsapras et al. 2019), and the second examined the population of trial solutions from the DE exploration of the FSBL model (R. A. Street et al. 2019).

For the grid search, we constructed a $\chi^2$ map spanning $\log s \in [-0.6, 0.6]$ in steps of 0.01 and $\log q \in [-5, -0.5]$ in steps of 0.05, giving 11,011 grid points. For each pair of ($\log s$, $\log q$), we performed a TRF fit using the TESS-only parameters from the earlier USBL modeling as initial guesses for $t_0$, $u_0$, $t_E$, $\rho$, and $\alpha$. The values of $\log s$ and $\log q$ were held fixed through narrow parameter bounds. The resulting grid showed a clear $\chi^2$ minimum at ($\log s$, $\log q$) $\approx$ (0.070, $-1.85$), and this location was adopted as the starting point for the subsequent MCMC run.

For the second approach, we investigated the likelihood landscape from the DE sampling performed during the FSBL stage. As described in the previous method, we examined the distribution of trial solutions and their associated $\chi^2$ values as a function of ($\log s$, $\log q$). The DE samples clustered around a

**Table 2**
Complete Microlensing Model Comparison Table

| | Gaia | | TESS S64 | | All TESS | Joint TESS+Gaia | | | |
|---|---|---|---|---|---|---|---|---|---|
| Parameter | PSPL | FSPL | PSBL | USBL | USBL | USBL | FSBL(main) | FSBL(alt) | FSBL($\pi$) |
| $t_0$ (JD 2.46E6) | $57.652^{+4.387}_{-2.298}$ | $57.542^{+3.865}_{-3.120}$ | $58.510^{+1.190}_{-0.350}$ | $58.994^{+0.035}_{-0.049}$ | $61.951^{+0.314}_{-0.184}$ | $66.981^{+0.642}_{-0.427}$ | $\mathbf{71.883}^{+0.508}_{-0.596}$ | $56.904^{+0.044}_{-0.040}$ | $70.670^{+0.738}_{-0.775}$ |
| $u_0$ | $0.000^{+0.008}_{-0.007}$ | $0.001^{+0.020}_{-0.014}$ | $0.042^{+0.001}_{-0.002}$ | $0.282^{+0.003}_{-0.003}$ | $0.310^{+0.003}_{-0.004}$ | $0.291^{+0.005}_{-0.004}$ | $\mathbf{0.260}^{+0.005}_{-0.004}$ | $0.072^{+0.002}_{-0.002}$ | $0.273^{+0.004}_{-0.005}$ |
| $t_E$ (days) | $981.600^{+4347.550}_{-692.300}$ | $396.670^{+681.140}_{-180.710}$ | $547.470^{+25.445}_{-23.700}$ | $25.543^{+0.547}_{-0.419}$, | $42.563^{+1.668}_{-1.009}$ | $82.205^{+4.483}_{-4.964}$ | $\mathbf{116.775}^{+3.891}_{-7.475}$ | $89.830^{+1.400}_{-1.650}$ | $82.930^{+2.080}_{-2.560}$ |
| $\rho$ ($10^{-3}$) | ⋯ | $18.830^{+19.480}_{-14.320}$ | ⋯ | $1.525^{+0.041}_{-0.049}$ | $0.898^{+0.038}_{-0.034}$ | $0.475^{+0.026}_{-0.024}$ | $\mathbf{0.338}^{+0.020}_{-0.012}$ | $0.281^{+0.013}_{-0.011}$ | $0.450^{+0.021}_{-0.021}$ |
| $s$ ($\theta_E$) | ⋯ | ⋯ | $1.094^{+0.002}_{-0.002}$ | $1.171^{+0.002}_{-0.003}$ | $1.163^{+0.002}_{-0.002}$ | $1.165^{+0.002}_{-0.003}$ | $\mathbf{1.153}^{+0.003}_{-0.002}$ | $0.953^{+0.001}_{-0.001}$ | $1.157^{+0.002}_{-0.004}$ |
| $q$ ($10^{-3}$) | ⋯ | ⋯ | $1.480^{+0.079}_{-2.570}$ | $61.782^{+2.711}_{-3.231}$ | $17.718^{+1.049}_{-1.450}$ | $4.312^{+0.626}_{-0.404}$ | $\mathbf{2.010}^{+0.269}_{-0.147}$ | $1.629^{+0.051}_{-0.044}$ | $3.620^{+0.280}_{-0.310}$ |
| $\alpha$ | ⋯ | ⋯ | $1.180^{+0.110}_{-0.080}$ | $5.332^{+0.013}_{-0.014}$ | $5.320^{+0.012}_{-0.009}$ | $1.053^{+0.013}_{-0.012}$ | $\mathbf{1.018}^{+0.014}_{-0.012}$ | $3.703^{+0.005}_{-0.006}$ | $0.908^{+0.017}_{-0.014}$ |
| $\pi_{E,E}$ | ⋯ | ⋯ | ⋯ | ⋯ | ⋯ | ⋯ | ⋯ | ⋯ | $0.494^{+0.004}_{-0.010}$ |
| $\pi_{E,N}$ | ⋯ | ⋯ | ⋯ | ⋯ | ⋯ | ⋯ | ⋯ | ⋯ | $-0.462^{+0.037}_{-0.025}$ |
| $f_{\rm S,Gaia}$ | 11.017 | 22.498 | ⋯ | ⋯ | ⋯ | 476.137 | **327.615** | 164.666 | 420.723 |
| $f_{\rm B,Gaia}$ | 1505.039 | 1496.100 | ⋯ | ⋯ | ⋯ | 1042.487 | **1188.523** | 1360.493 | 1102.842 |
| $G_{\rm S,Gaia}$ | 24.793 | 24.019 | ⋯ | ⋯ | ⋯ | 20.705 | **21.110** | 21.860 | 20.840 |
| $G_{\rm B,Gaia}$ | 19.457 | 19.463 | ⋯ | ⋯ | ⋯ | 19.855 | **19.710** | 19.570 | 19.790 |
| $f_{\rm S,TESS}$ | ⋯ | ⋯ | 1917.658 | 9348.192 | 10,659.916 | 12,238.181 | **12,626.561** | 2738.681 | 9613.809 |
| $f_{\rm B,TESS}$ | ⋯ | ⋯ | 591,016.517 | 597,887.784 | 598,267.332 | 590,379.953 | **584,911.045** | 605,702.108 | 596,197.042 |
| $\chi^2$ | 51.60 | 51.58 | 150114.58 | 66124.47 | 112633.35 | 109143.28 | **106811.33** | 135078.80 | 242589.13 |
| $\chi^2_\nu$ | 0.379 | 0.379 | 13.220 | 5.551 | 5.468 | 5.264 | **5.151** | 6.515 | 11.700 |

**Note.** Columns are grouped by dataset (Gaia, TESS Sector 64, All TESS, and Joint TESS+Gaia). All values are medians with 16th/84th percentile uncertainties. Gaia magnitudes are computed via Equation (11) (M. Riello et al. 2021).

different location, near $(\log s, \log q) \approx (-0.062, -2.35)$, suggesting a second minimum. Both regions therefore merited full exploration. Both $\chi^2$ analysis figures can be found in Appendix A.

We investigated both solutions independently through FSBL MCMC fits. To avoid biasing the search toward only one branch of the classic close–wide degeneracy, we imposed simple limits on the separation parameter, in one case setting $\log s > 0$ and in a second case setting $\log s < 0$. We kept the mass-ratio limits broad ($\log q \in [-3, -1]$), so both candidate solutions were fully enclosed. The solution near $(0.070, -1.85)$ consistently converged to the model adopted in Table 2 and provided the lowest overall $\chi^2$. A corner plot for this solution is shown in Figure 3.

The solution near $(-0.062, -2.35)$ also converged, although its likelihood peak was strongly influenced by a sharp feature near JD $\approx$ 2460040, which coincides with the Sector 63–64 boundary. The Data Release Notes for both sectors show background flux ramps at the ends and beginnings of the sectors (M. M. Fausnaugh et al. 2023a, 2023b), consistent with scattered-light-driven changes in the spacecraft background rather than a physical microlensing signal. Such short-timescale ramps and similar artifacts often appear in TESS light curves. As the global $\chi^2$ of this model is higher than that of the adopted solution, we determined that this solution was most likely a result of the TESS systematics.

The parameters, corner plot, caustic geometry, and model light curves for the second solution are shown in Appendix B, and this solution is included in Table 2 for completeness, but we do not use this model in the physical interpretation.

#### 3.4.2. Parallax

Given the spatial separation between TESS (in elliptical Earth orbit) and Gaia (at L2), the combined light curves offer a potential baseline for constraining the microlensing parallax. As such, we attempted an additional fit for a full-parallax FSBL model, which adds the north ($\pi_{EN}$) and east $\pi_{\rm EE}$ components of the microlensing parallax to our parameter list. We used our static FSBL model solution as an initial guess for a quick TRF fit, with $\pi_{EN} = \pi_{EE} = 0.005$ to begin, followed by a long MCMC fit to better explore this new configuration.

The parallax model formally converged to $\pi_{{\rm E},N} = -0.145 \pm 0.010$ and $\pi_{{\rm E},E} = 4.768 \pm 0.051$. From these values, we inferred the corresponding physical parameters—with Equation (8), a relative parallax of $\pi_{\rm rel} \approx 0.645$ mas, which places the lens at $D_L \approx 0.2$ kpc, assuming a source star within the the Milky Way ($D_S \approx$ 11–13 kpc). A lens at this distance with a mass near the `pyLIMASS` posterior ($M_L \approx 0.8\,M_\odot$) would have $G \approx 13.4$ after extinction (based on our calculation described in Section 4.2), which is substantially brighter than the blended flux inferred from the joint fit ($G_{\rm blend} \approx 19.7$). This inconsistency indicates that the parallax solution is not physically viable.

We therefore examined the posterior only for the restricted parallax fit, in which both $\pi_{{\rm E},N}$ and $\pi_{{\rm E},E}$ were limited to a physically reasonable range for Galactic microlensing ($\pi_{EN} \in [0,\ 0.5]\pi_{{\rm E},E} \in [0,\ 0.5]$). In this case, the MCMC chains drove both $\pi_{{\rm E},N}$ and $\pi_{{\rm E},E}$ to the edges of the allowed interval, producing narrow, non-Gaussian posteriors that are dominated by the prior bounds rather than the data. Systematic effects in the light curves—particularly those associated with TESS background variations and sector boundaries—may further contribute to spurious parallax signals and complicate the interpretation of the fit. Given these factors, we do not adopt the parallax solution and instead favor the static model as the most physically consistent description of the event.

For completeness, we include the same information for the parallax as for the alternate static solution in Appendix C, and this solution is included in Table 2.

In the following results section, we adopt the parameters from the joint Gaia+TESS FSBL configuration without parallax as our primary solution for describing the system, which achieved the best reduced $\chi^2_\nu$ value, where $\chi^2_\nu = \chi^2/N_{\rm data}$. The photometric observations and best-fit model are shown in Figure 2, the posterior distributions for key parameters are presented in Figure 3, and the predicted source trajectory and caustic geometry are illustrated in Figure 4. This approach incorporates both datasets, provides a strong baseline measurement, delivers the best overall $\chi^2$ fit, and takes advantage of the complementary coverage of Gaia and TESS to reduce parameter degeneracies and improve the precision of the derived quantities.

## 4. Results

### 4.1. Source Star Parameters

The source star for the Gaia23bra event corresponds to Gaia DR3 5252141822116891648—a $G = 19.38$ mag star located along the Galactic Plane. As this star is fainter than $G = 19$ mag, its stellar parameters were not included in the Gaia DR3 Apsis II release (M. Fouesneau et al. 2023). Additionally, the Gaia parallax measurement is negative, implying an unphysical distance and rendering it unusable for direct inference. As most microlensing events are located toward the Galactic Bulge, the Galactic population models generally applied to estimate source and lens properties are designed to statistically represent the stellar populations in that direction. They are, however, less reliable in describing populations located along the Galactic Plane. We therefore applied the available photometric and astrometric data from Gaia DR3, combined with parameters from our microlensing model, to the Gaussian mixture inference algorithm in `pyLIMASS` (E. Bachelet et al. 2024) to characterize the source.

`pyLIMASS` applies a multidimensional Gaussian mixture model to combine photometry, astrometry, stellar isochrones, and the angular source size $\rho$ and microlensing timescale $t_{\rm E}$ from our `pyLIMA` microlensing model. Based on the hypothesis discussed in Section 3.1.1, for the photometry of the source star, we used the modeled source flux from Equation (10) as the photometric input to `pyLIMASS`, rather than the total blended Gaia magnitude. With this flux, the `pyLIMASS` analysis yielded a source star mass of $M_{\rm S} = 0.897^{+0.154}_{-0.151} M_\odot$, and a distance $D_{\rm S} = 12.860^{+1.419}_{-2.102}$ kpc. We note that the posterior distribution for $D_{\rm S}$ is skewed toward larger values, which is common in `pyLIMASS` analyses when Gaia parallaxes are unreliable. As noted in E. Bachelet et al. (2024), this algorithm tends to overestimate source distances, as its Galactic priors are tuned for Bulge events, which can bias sources in the Plane. This explains the asymmetry seen in the lower left image of Figure 5. The final source star parameters are presented in Table 3.

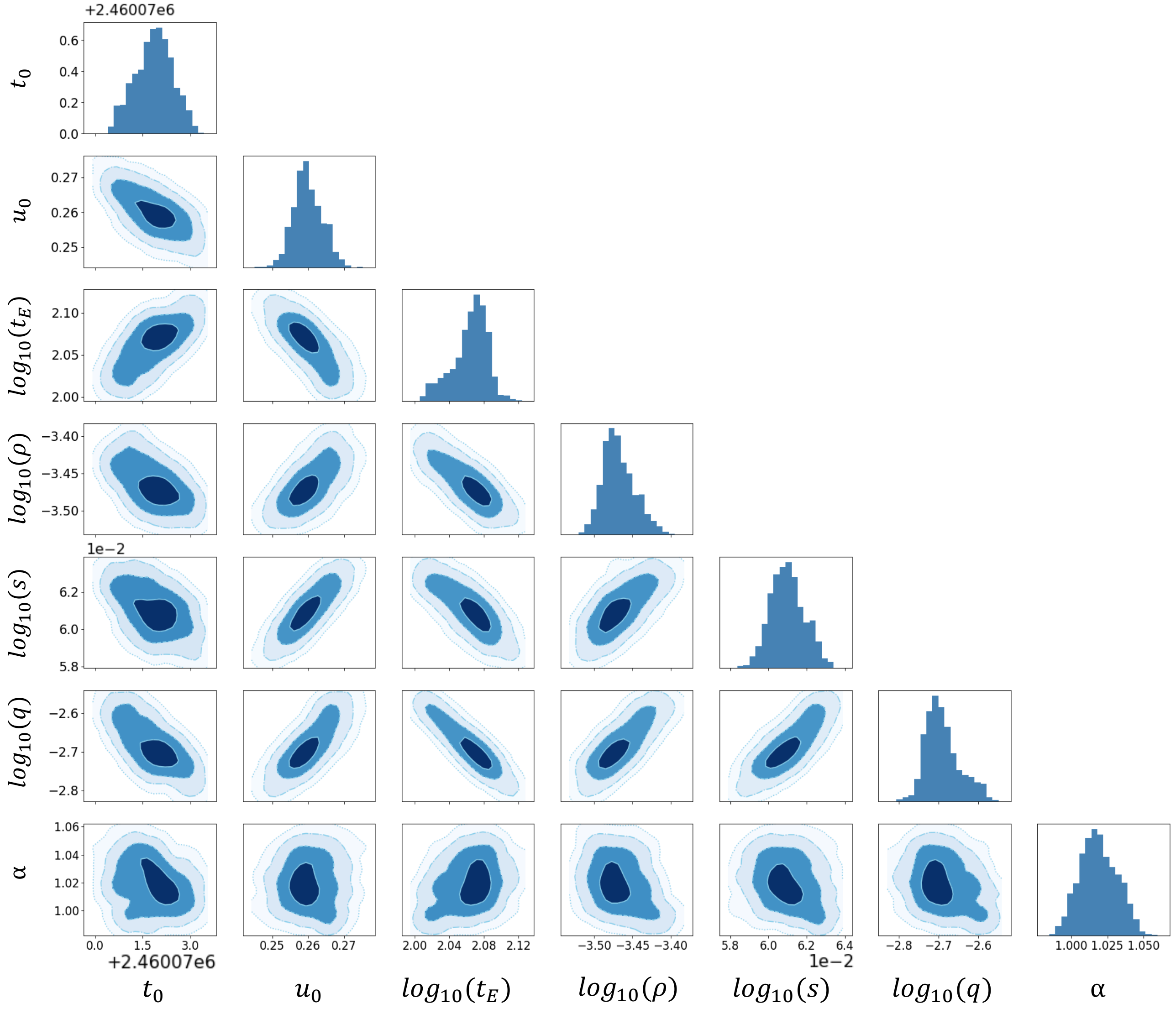

**Figure 3.** Corner plot of posterior distributions for the key model parameters of the `pyLIMA` fit depicted in Figure 2, including $t_0$, $u_0$, $\log t_{\rm E}$, $\log \rho$, log separation, log mass ratio, and $\alpha$. The contours indicate the $1\sigma$–$4\sigma$ credible regions of the posterior distributions.

Its Galactic coordinates at $l \approx 287°$, $b \approx -5°$, paired with this distance of $\sim$12 kpc, places the source star at $\sim$1 kpc below the Galactic Plane (G. Gilmore & N. Reid 1983; K. Vieira et al. 2022). Its Gaia proper motions (see Table 1) imply a tangential velocity of approximately 188 km s$^{-1}$, further supporting an association with the older disk population. These characteristics indicate that the source is a late G dwarf located in the thick disk.

### 4.2. Lens Star Parameters

As demonstrated in Section 3.1, the mass and distance to the lens star are crucial for determining the nature of a microlensing binary. The lens mass $M_L$ and the angular Einstein radius $\theta_E$ serve as the scaling parameters for interpreting the microlensing companion—specifically, for transforming the mass ratio $q$ and the normalized projected separation $s$ to the true planetary mass and semimajor axis. If the distances to both the lens and source stars are known, the relative parallax $\pi_{\rm rel}$ (Equation (7)) can be computed. Using this value and the lens mass to evaluate $\theta_E$ (Equation (6)), we can derive the physical separation and mass of the companion, fully characterizing the binary system.

We estimated the lens mass using the same `pyLIMASS` analysis applied to the source star. From the posterior distribution, we inferred a distance of $D_L = 4.330^{+1.003}_{-1.012}$ kpc and a mass of $M_L = 0.774^{+0.170}_{-0.146}\,M_\odot$, consistent with an early K dwarf star. Using these posteriors, we applied Equation (6) to calculate $\theta_E = 0.977^{+0.115}_{-0.100}$ mas.

Using the lens mass posterior from the `pyLIMASS` analysis, we estimated the intrinsic luminosity of the lens star using the mass–luminosity relations of M. Cuntz & Z. Wang (2018). We then applied bolometric corrections from the Gaia DR2 calibration (R. Andrae et al. 2018). Line-of-sight extinction was incorporated using the $A_V$ posterior inferred by `pyLIMASS` and converted from the Johnson–Cousins $V$ band to the $G$ band using the Gaia EDR3 photutils module (F. Anders et al. 2022). This procedure yields an apparent lens magnitude of $G_{\rm lens} = 20.06^{+1.69}_{-1.38}$. This value is statistically consistent with the blended light inferred from the microlensing fit ($G_{\rm blend} = 19.71$), indicating that the lens could account for a significant fraction, and potentially the entirety, of the blended light in the Gaia observations. Given the spatial resolution of the available data, however, it is not possible to distinguish between a scenario in which the lens dominates the blend and one in which additional unrelated sources contribute to the observed flux. Future high-resolution imaging will be required to resolve the source and lens and determine the origin of the blended light.

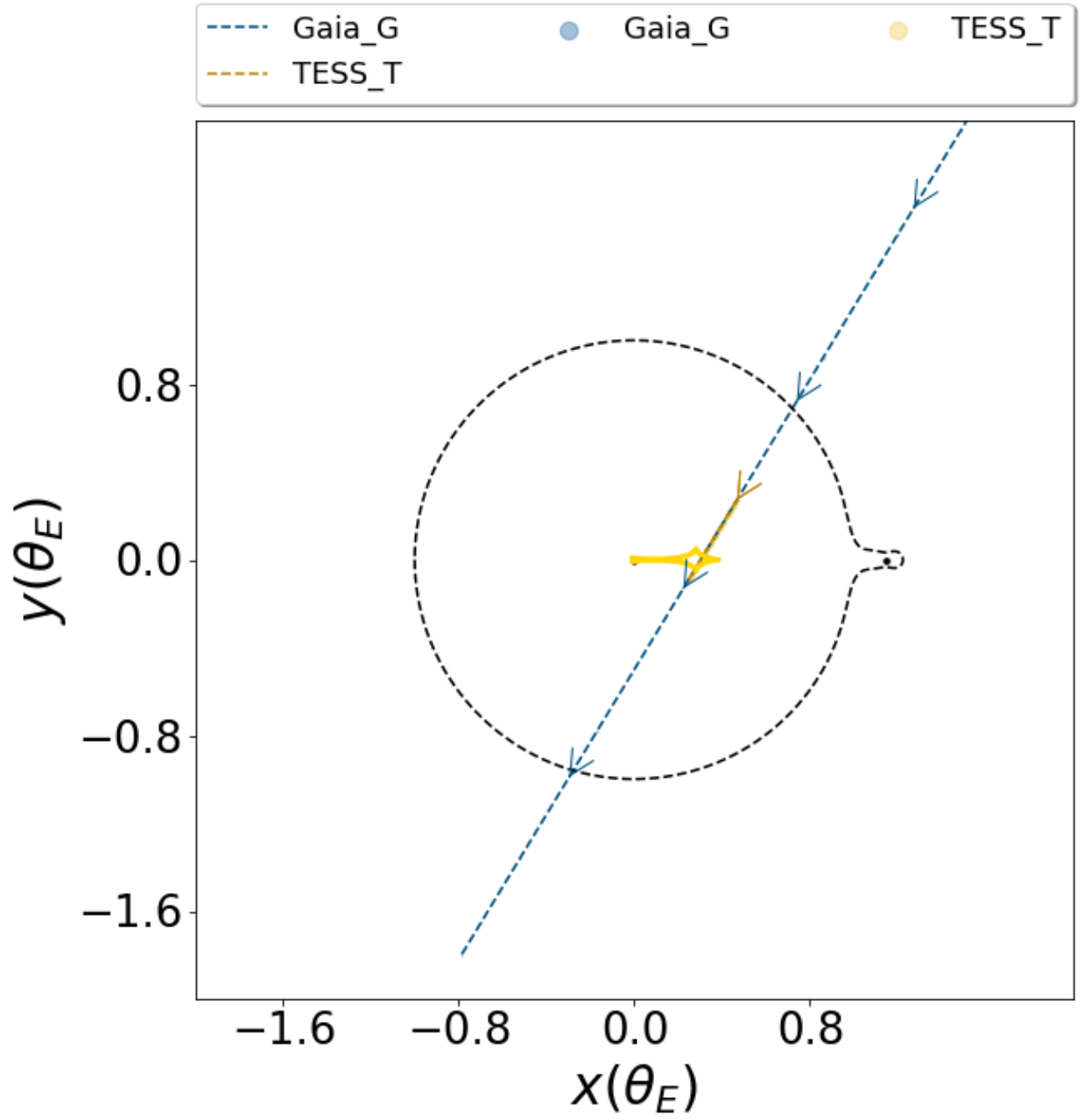

**Figure 4.** Source trajectory in the lens reference frame, showing the caustic structure (yellow) and the path of the source relative to the lens system as observed by Gaia (blue) and TESS (gold), based on the pyLIMA model depicted in Figure 2 (E. Bachelet et al. 2017). The arrows indicate the direction of motion.

To assess the feasibility of current instruments resolving this system with future observations, we considered the relative motion between the source and lens. From our pyLIMASS analysis, we derived a relative proper motion between the lens and source of $\mu_{\rm rel} = 3.084^{+0.391}_{-0.340}$ mas yr$^{-1}$. With this relative velocity, the lens and source will separate by approximately 15 mas in 5 yr and ∼30 mas in 10 yr.

This level of separation is resolvable with current space-based facilities such as the Hubble Space Telescope or with ground-based adaptive optics on large telescopes, if the lens is sufficiently bright. We therefore estimated the lens apparent magnitude in the H/K bands to assist in these follow-up observations, following the same method we applied previously to estimate the brightness in the G band. For the H/K magnitudes, we neglect bolometric corrections (L. Casagrande & D. A. VandenBerg 2014), as their variation in the near-infrared is small for late-type dwarfs ($\Delta BC_{H,K_s} \lesssim 0.07$), and thus does not contribute significant additional uncertainty. We thus obtain apparent magnitudes of $m_K \approx 17.90^{+0.45}_{-0.57}$ and $m_H \approx 18.00^{+0.46}_{-0.58}$ (M. Cuntz & Z. Wang 2018; C. N. A. Willmer 2018). As the Hubble limiting magnitude in 1 hr exposure ranges over $H_{\rm mag} \sim 25.9$–28.4, the lens star is well within Hubble's ability to observe, making future high-resolution imaging a viable way to confirm the lens properties and fully solve the system. All these magnitudes and lens star properties are presented in Table 3.

Without a direct parallax measurement, the lens mass remains model-dependent and cannot be definitively classified. For the purposes of this analysis, we adopt the inferred mass and its associated uncertainties when deriving the physical properties of the companion.

### 4.3. Planetary Properties

Utilizing the joint FSBL configuration, we measured the projected separation parameter as $s = 1.153^{+0.003}_{-0.002}$ and the mass ratio as $q = (2.01^{+0.27}_{-0.15}) \times 10^{-3}$. The projected separation, calculated using the inferred Einstein radius and lens distance derived in the previous section, yields $a_{\perp,\rm min} = 4.85^{+0.77}_{-0.81}$ AU. Using this projected separation, the companion lies at a distance comparable to Jupiter's orbit, consistent with the population of cold gas giants typically probed by microlensing. As $a_{\perp,\rm min}$ represents the instantaneous projected separation, rather than the true semimajor axis, this value serves only as a lower limit on the orbital distance.

The measured mass ratio falls below the planetary threshold of $q_{\rm max} = 0.03$ (D. Suzuki et al. 2016), classifying the secondary object as a planetary companion of the lens star. Combined with the lens mass from Section 4.2, we estimate the mass of this object to be $M_P = 1.63^{+0.42}_{-0.38}\, M_{\rm Jup}$. This mass places the companion in the super-Jupiter regime.

We thus announce Gaia23bra b as the first gravitationally bound microlensing planet discovered by TESS.

## 5. Discussion

### 5.1. Microlensing Capabilities of TESS

Previous yield estimates suggested TESS would rarely detect microlensing planets (H. Yang et al. 2024; M. Kunimoto et al. 2025), but Gaia23bra b demonstrates that when combined with long-baseline observations, TESS can provide meaningful contributions to microlensing science. While not designed for this purpose, TESS's unparalleled 200 s cadence —introduced in Extended Mission 2—provided dense coverage of caustic-crossing features of the Gaia23bra event that were essential for constraining the mass ratio and projected separation of the binary lens.

#### 5.1.1. Large Pixel Scale and Source Identification

TESS's coarse spatial resolution is one of its most significant limitations for microlensing science. With 21” pixels, the source star cannot be distinguished from its neighbors, and blending from other stars in the pixel introduces substantial background flux. In the case of Gaia23bra b, Gaia's long-baseline observations allowed us to identify the source independently, ensuring that the microlensing signal extracted from TESS could be confidently associated with the correct star. Future detections may similarly benefit from external astrometric catalogs or high-resolution imaging to identify the source star and mitigate blending.

#### 5.1.2. Short Observing Windows and Event Coverage

TESS's 27 day sectors further limit its ability to capture full microlensing baselines, which are essential for constraining the Einstein timescale and, by extension, lens mass and distance. This constraint introduces a natural bias: only events with short-duration, high-magnification features are likely to be detected. Gaia23bra illustrates this point—the caustic crossings occurred entirely within a single sector, allowing us to model the binary-lens configuration despite missing the full ingress and egress. Although TESS cannot replace long-baseline surveys, its cadence enables the precise characterization of short-lived features that are essential for detecting planetary companions.

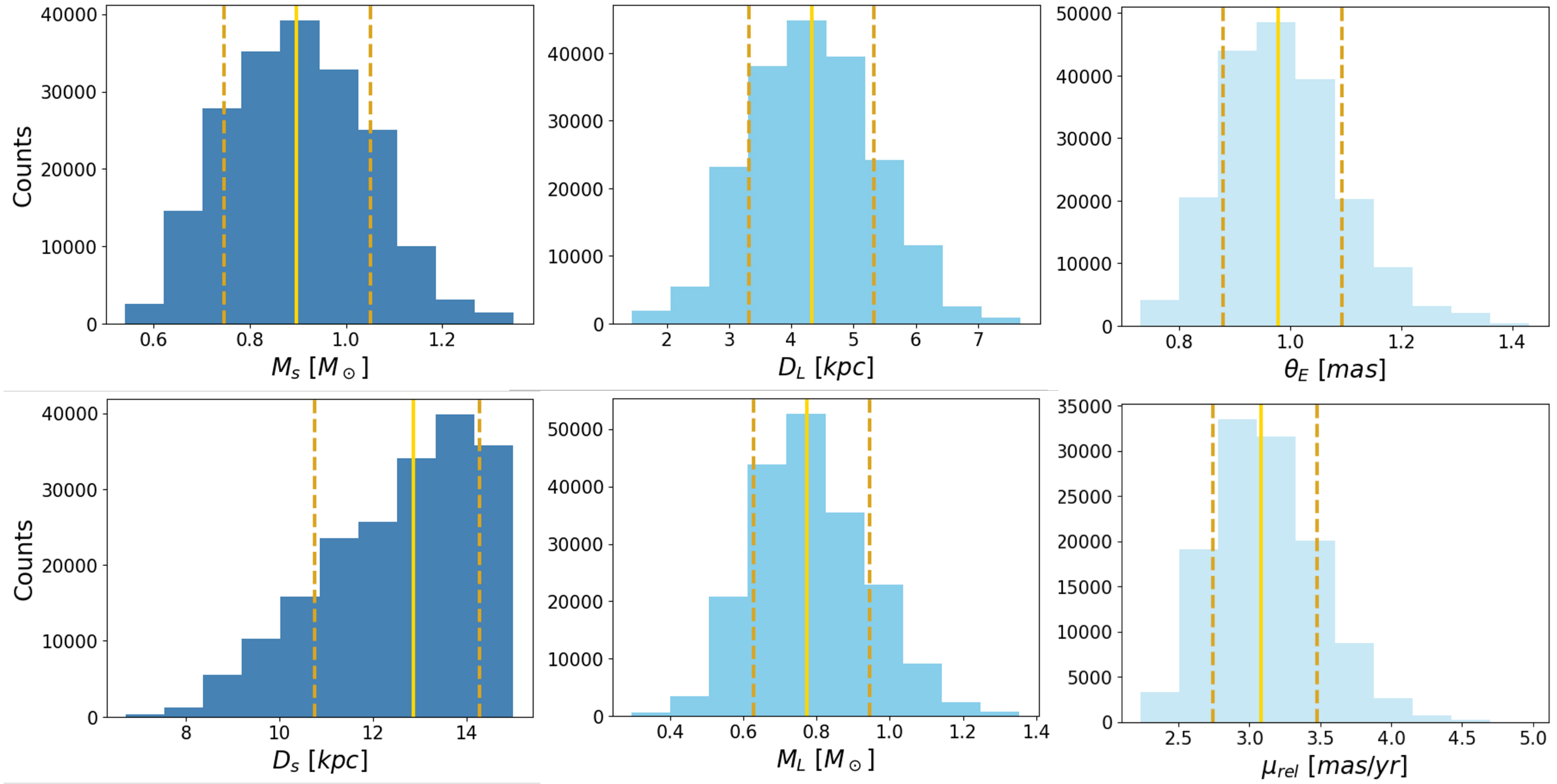

**Figure 5.** Posterior distributions from the `pyLIMASS` (E. Bachelet et al. 2024) mass estimation run, showing histograms for key lens and source parameters: the source distance ($D_S$) and source mass ($M_S$) (dark blue), the lens distance ($D_L$) and lens mass ($M_L$) (bright blue), and the Einstein radius ($\theta_E$) and relative proper motion ($\mu_{rel}$) (pale blue). The solid gold vertical lines indicate the posterior mean values, while the dashed dark gold lines mark the 16th and 84th percentiles.

Due to the high background levels in TESS full-frame images, only the brightest portion of a microlensing event is typically detectable, causing the observed light curve to miss the low-amplitude rise and decline. This truncation shortens the apparent event timescale recovered from pixel lensing, and in cases like Gaia23bra, the $t_E$ inferred from TESS alone is substantially smaller than the true value measured from long-baseline data. This behavior is expected in pixel-lensing regimes and underscores the need for complementary observations, such as Gaia, to anchor the full light-curve shape.

Consequently, only high-magnification or binary-lens events with short-duration features are likely to be detectable in TESS photometry, as these produce sharp, high-contrast signatures even in the absence of baseline flux measurements. These events therefore represent the most promising microlensing targets for TESS.

**Table 3**
Estimates for Stellar Properties for the Source and Lens Star Using `pyLIMASS` and the Final Estimates for the Mass and Projected Semimajor Axis of the Planetary Companion

| Parameter | Value | Source |
|---|---|---|
| $M_S$ ($M_\odot$) | $0.897^{+0.154}_{-0.151}$ | `pyLIMASS` |
| $D_S$ (kpc) | $12.860^{+1.419}_{-2.102}$ | `pyLIMASS` |
| $m_S(G_{mag})$ | $21.34^{+1.14}_{-1.68}$ | `pyLIMASS+pyLIMA` |
| $M_L$ ($M_\odot$) | $0.774^{+0.170}_{-0.146}$ | `pyLIMASS` |
| $D_L$ (kpc) | $4.330^{+1.003}_{-1.012}$ | `pyLIMASS` |
| $\mu_{rel}$ (mas yr$^{-1}$) | $3.084^{+0.391}_{-0.340}$ | `pyLIMASS` |
| $\theta_E$ (mas) | $0.977^{+0.115}_{-0.100}$ | `pyLIMASS+pyLIMA` |
| $m_L(G_{mag})$ | $20.06^{+1.69}_{-1.38}$ | `pyLIMASS+pyLIMA` |
| $m_L$ ($H_{mag}$) | $18.00^{+0.46}_{-0.58}$ | `pyLIMASS+pyLIMA` |
| $m_L$ ($K_{s,mag}$) | $17.90^{+0.45}_{-0.57}$ | `pyLIMASS+pyLIMA` |
| $a_{\perp,min}$ (AU) | $4.847^{+0.770}_{-0.808}$ | `pyLIMASS+pyLIMA` |
| $M_P$ ($M_{Jup}$) | $1.630^{+0.420}_{-0.378}$ | `pyLIMASS+pyLIMA` |

### *5.1.3. High Cadence and Temporal Resolution*

TESS's 200 s cadence provides a clear and continuous view of Gaia23bra and was essential for identifying the caustic-crossing structure that motivated the binary-lens interpretation. Even when binned to longer cadences, the TESS observations remain significantly denser than Gaia and typical ground-based microlensing surveys, which often sample on day- to month-long timescales (T. Sumi et al. 2003; A. Udalski 2003; Gaia Collaboration et al. 2016; D.-J. Kim et al. 2018).

To illustrate the broader relevance of this detection, we also examined how the light curve appears when binned to 12 minute intervals, to approximate Roman's planned sampling. The crossings remain detectable at this cadence (see Figure 6), and the overall morphology of the event is preserved. This comparison demonstrates how similar events to Gaia23bra could be identifiable in Roman's survey data, and highlights the advantage of high-cadence, space-based monitoring for revealing short-timescale features that would be missed by surveys sampling at ∼one point per night or less (T. Sumi et al. 2003; A. Udalski 2003; D.-J. Kim et al. 2018).

Finite-source effects play a central role in breaking microlensing degeneracies, because the detailed morphology of caustic-crossing features encodes the source size and trajectory. These signatures tighten constraints on $q$, $s$, $\rho$, and $\alpha$, and can only be recovered when the light curve is sampled at sufficiently high cadence.

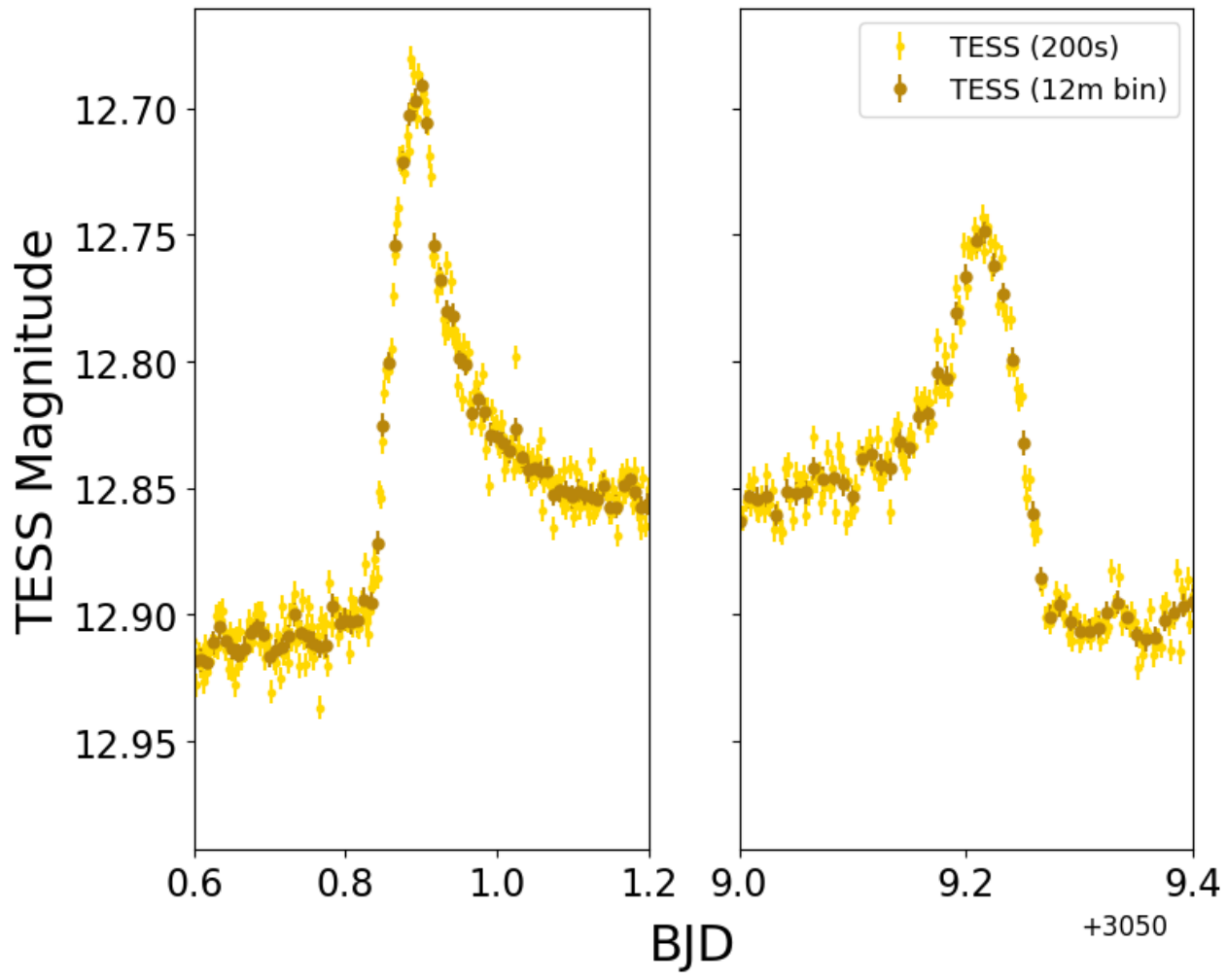

**Figure 6.** Zoomed view of the TESS caustic-crossing events with a cadence of 200 s (gold) and a binned cadence of 12 minutes (dark gold) chosen to approximate Roman's planned sampling. As shown, the 12 minute cadence provides only two to three points near the peaks, while the TESS observations provide roughly 5 times as many. This illustrates that space-based surveys with minute-level cadence can reliably detect caustic crossings, and that higher-cadence sampling further reveals the detailed morphology and finite-source curvature that improve physical constraints.

While hour-level sampling is sufficient to recover the overall morphology of multihour caustic crossings, higher-cadence observations provide significantly improved resolution of the detailed curvature and finite-source structure, enabling tighter constraints on the physical parameters when fully leveraged.

This result also follows the precedent set by earlier space-based microlensing programs, including Kepler's K2 Campaign 9 (A. Gould & K. Horne 2013; S. B. Howell et al. 2014; C. B. Henderson et al. 2016) and Spitzer's parallax measurements (S. Dong et al. 2007; A. Udalski et al. 2015), both of which demonstrated that general-purpose photometric missions can be repurposed to deliver key microlensing constraints.

High-cadence coverage not only enabled the identification of the finite-source structure, but also strengthened the constraints on parameters tied to the event geometry. Coarser temporal sampling weakens these constraints and makes quantities such as the mass ratio, projected separation, and source size more difficult to disentangle. The short-lived structure of the caustic crossings also helps refine the impact parameter and event timescale; both become less precisely determined when key portions of the light curve are sparsely sampled.

Taken together, these comparisons show that while TESS's cadence was critical for the detailed characterization of Gaia23bra, similar events would still be detectable at Roman's planned sampling. The Gaia23bra event therefore highlights how the combination of high-cadence, space-based photometry and long-baseline monitoring provides a powerful approach for identifying and characterizing short-timescale microlensing features, and suggests that TESS may continue to contribute to the growing sample of microlensing exoplanets, particularly in regions outside the Galactic Bulge.

### 5.2. Galactic Plane Microlensing

In addition to being TESS's first definitive microlensing detection, the Gaia23bra event is notable for its location far from the Galactic Bulge, where most microlensing surveys have traditionally focused. Recent studies using Gaia, the Optical Gravitational Lensing Experiment (OGLE) IV, and the Zwicky Transient Facility have shown that microlensing events occur along the Galactic Plane, though at lower expected yields compared to the Bulge (P. Mróz et al. 2020; A. C. Rodriguez et al. 2022; Ł. Wyrzykowski et al. 2023). For TESS, this represents an opportunity rather than a challenge: its all-sky survey provides "free" coverage of these regions without dedicated observing time, enabling the exploration of planetary populations in diverse galactic environments.

Gaia23bra b is one of only a handful of microlensing planets discovered in the Galactic Plane, and its location near $l \sim 280°$ places it in a region where multiple other candidate microlensing events have also been identified. Notably, the planetary-mass lens from Z. Wu et al. (2024) and the Gaia23bey event (M. Harris et al. 2026, in preparation) are both located near this same galactic longitude. P. Mróz et al. (2020) reported an excess of microlensing events near $l \sim 280°$, but concluded that the optical depth alone—despite this region's proximity to the tangent point of the Carina spiral arm—could not account for the increased event rate. The apparent clustering of microlensing events in this region remains unexplained and warrants further investigation.

The detection of Gaia23bra b adds to this growing body of evidence and underscores the potential of the Galactic Plane as a complementary region for microlensing surveys. Continued monitoring of this area may help reveal the cause of the event clustering and increase the number of microlensing planet detections. With its high-cadence coverage, TESS is an ideal instrument to initiate and support such efforts.

### 5.3. Future Outlook for Space-based Microlensing

Without a measurable parallax, the lens mass and distance remain model-dependent for the Gaia23bra system, leaving our characterization as precise in relative terms but incomplete in absolute physical properties. For the Bulge, future missions like Roman will improve these limitations through continuous coverage, high spatial resolution, and, when paired with simultaneous ground-based or other space-based observations, parallax baselines. Roman's orbit at L2 will introduce some parallax motion. Even so, simultaneous observations from a second observatory will still be essential for robust mass and distance determinations. For the Galactic Bulge, this can be achieved through ground-based microlensing surveys including the Microlensing Observations in Astrophysics (MOA), the Optical Gravitational Lensing Experiment (OGLE), and the Korea Microlensing Telescope Network (KMTNet).

Looking ahead, while the Gaia mission ended in 2025 January, the Vera C. Rubin Observatory's Legacy Survey of Space and Time (LSST) "Wide Fast Deep" coverage and its Dusty Plane minisurvey will provide long-baseline, higher-resolution observations that may complement TESS's high-cadence data (Vera C. Rubin Observatory LSST Solar System Science Collaboration et al. 2021). These capabilities will create new opportunities for joint analyses and extend TESS's contributions to microlensing science. Rubin's depth and spatial resolution will also aid source identification in crowded Plane fields, helping to mitigate the blending challenges inherent to TESS full-frame images.

## Acknowledgments

We are grateful to the referee for the constructive feedback, which significantly improved the manuscript.

M.H. and D.D. acknowledge support from the TESS Guest Investigator Program (80NSSC23K0769). This work was partially supported by an IPAC Visiting Graduate Student Research Fellowship.

This Letter includes data collected by the TESS mission, which are publicly available from the Mikulski Archive for Space Telescopes (MAST). Funding for the TESS mission is provided by NASA's Science Mission directorate. Some of the data presented in this Letter were obtained from the Mikulski Archive for Space Telescopes (MAST) at the Space Telescope Science Institute. The specific observations analyzed can be accessed via doi:10.17909/t9-9j8c-7d30, (M. Fausnaugh 2021).

We would like to thank the UNM Center for Advanced Research Computing, supported in part by the National Science Foundation, for providing the high performance computing resources used in this work.

This work has made use of data from the European Space Agency (ESA) mission Gaia (https://www.cosmos.esa.int/gaia), processed by the Gaia Data Processing and Analysis Consortium (DPAC; https://www.cosmos.esa.int/web/gaia/dpac/consortium). Funding for the DPAC has been provided by national institutions, in particular the institutions participating in the Gaia Multilateral Agreement. We also acknowledge the Gaia Photometric Science Alerts Team (http://gsaweb.ast.cam.ac.uk/alerts).

This research has made use of the VizieR catalog access tool, CDS, Strasbourg, France (F. Ochsenbein 1996). The original description of the VizieR service was published in F. Ochsenbein et al. (2000).

The authors acknowledge the assistance of Microsoft Copilot, powered by the GPT-5 model, in drafting and proofreading portions of this manuscript. Copilot was used to refine language, improve clarity, and ensure consistency in formatting, while all scientific content and interpretations remain the responsibility of the authors.

*Software:* pyLIMA (E. Bachelet et al. 2017), pyLIMASS (E. Bachelet et al. 2024), Microsoft Copilot (Microsoft 2025), Vizier (F. Ochsenbein et al. 2000).

## Appendix A Binary-lens $\Delta\chi^2$ Maps

This appendix presents the binary-lens $\Delta\chi^2$ maps used to identify candidate minima, shown in Figure A1. The left panel shows the fixed-grid $\Delta\chi^2$ map, while the right panel shows the DE-sampled likelihood landscape.

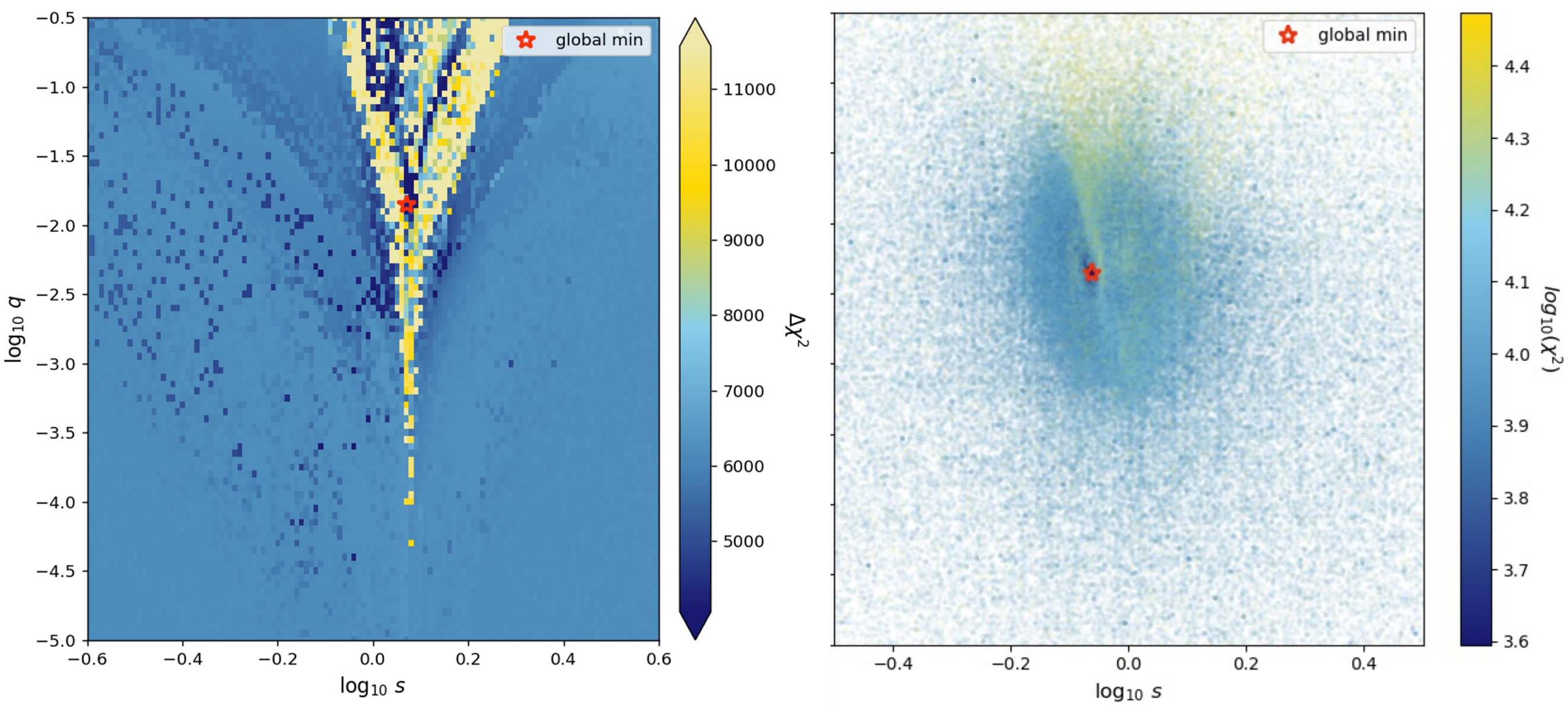

**Figure A1.** $\Delta\chi^2$ mapping of the binary-lens parameter space used to identify candidate minima in ($\log s$, $\log q$). Left: a grid in which each point corresponds to a fixed ($\log s$, $\log q$) pair for which the remaining FSBL parameters were optimized. The red star represents the location of the global minimum found. Right: a scatter plot depicting the population explored by the DE fit and the resulting $\chi^2$ for each solution. The darker blue colors denote lower $\chi^2$ values and better fits, with the red star denoting the global minimum. These are two low-$\chi^2$ regions discussed in Section 3.4.1.

## Appendix B
## Alternate Binary-lens Solution

This solution is documented for completeness but is not used in the physical interpretation (see Section 3.4.1). We include the light curve with this model overlayed, (Figure B1) , the geometry predicted fion the lens reference frame (Figure B2), and the corner plot showing the posteriors from the unfavored fit (Figure B3).

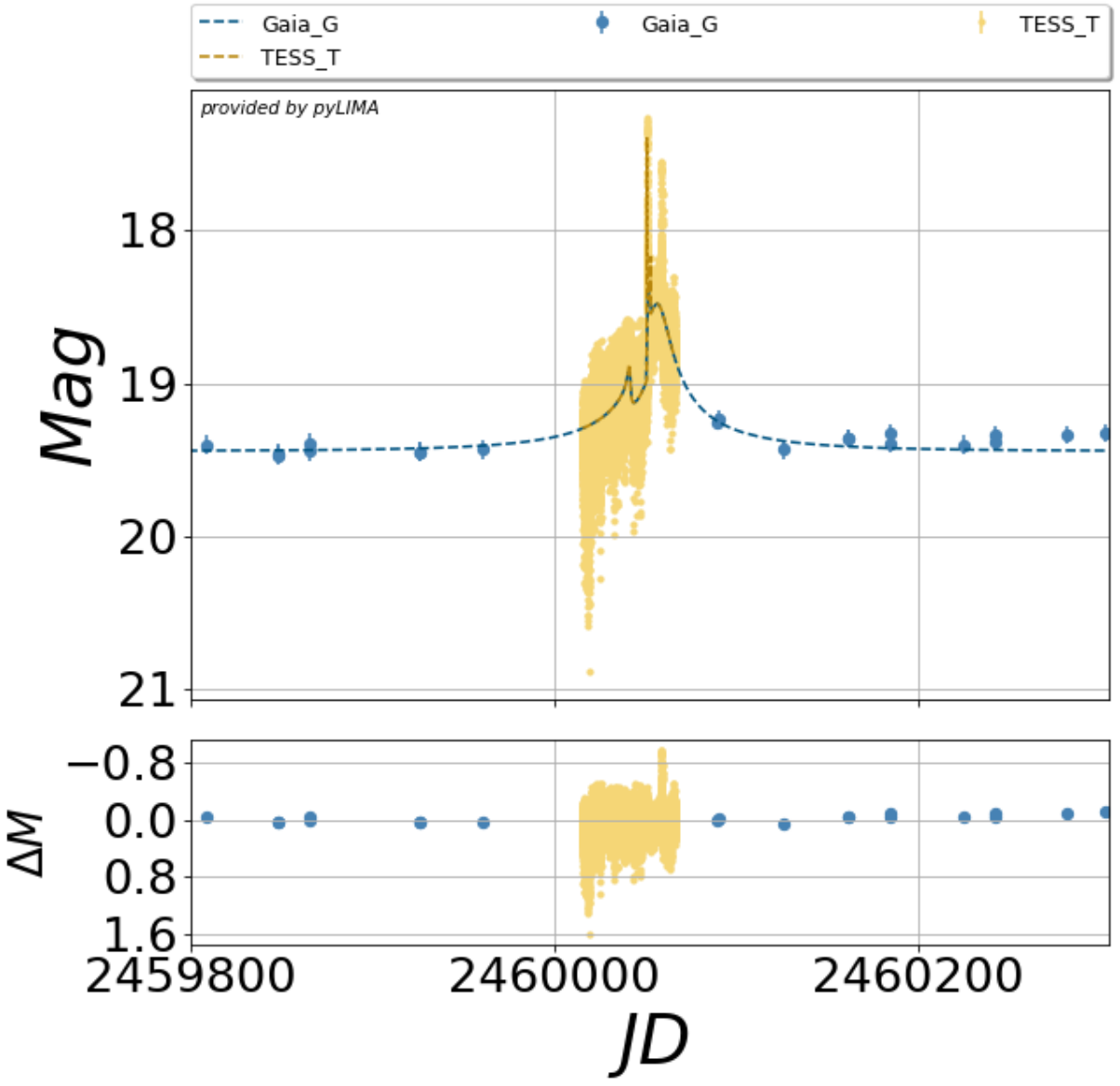

**Figure B1.** The photometry with the model overlaid for the alternate solution. The peak at the TESS sector break is well modeled, while the second caustic crossing is neglected. The format is identical to Figure 2.

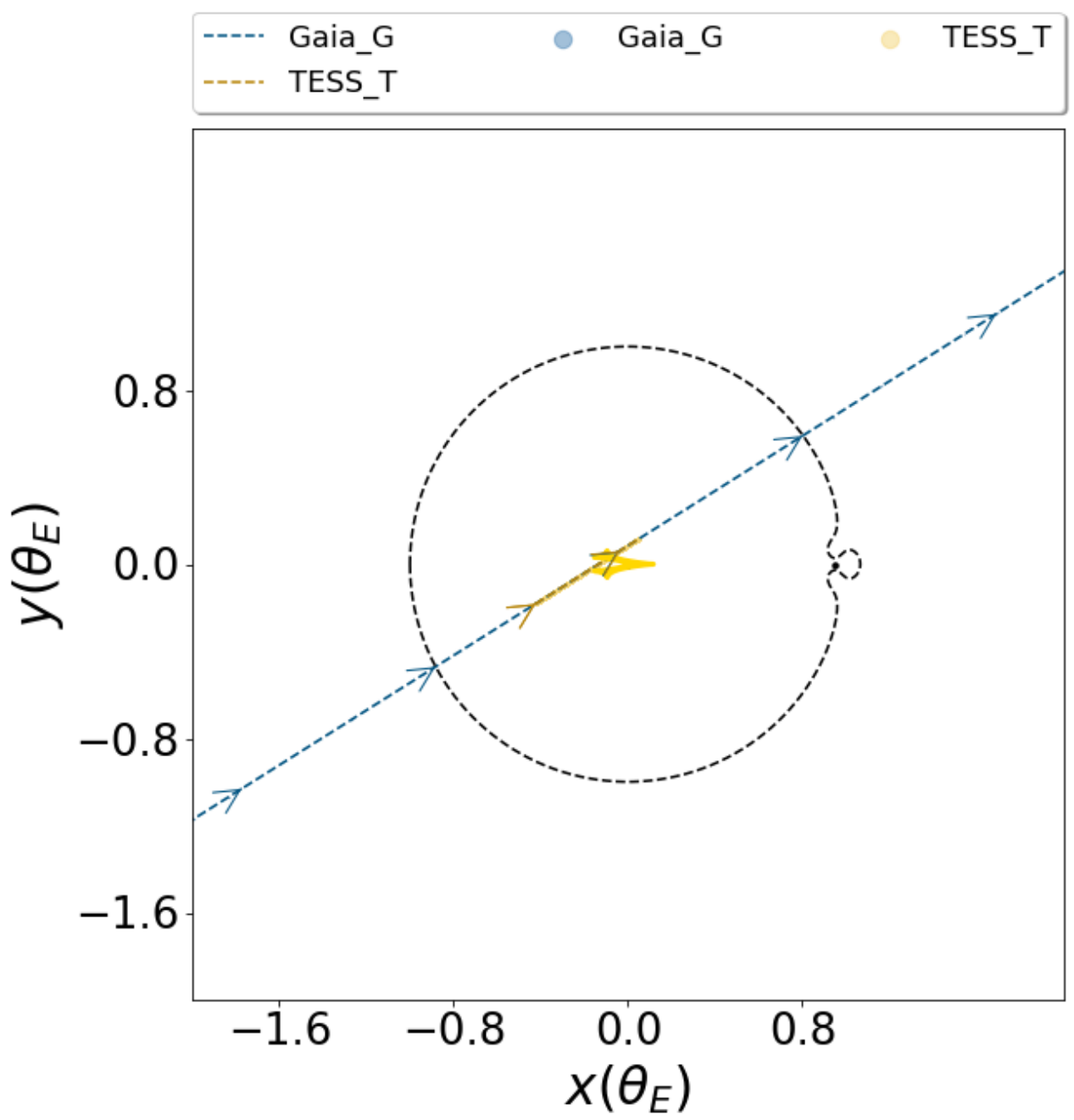

**Figure B2.** Geometry of the system from pyLIMA for the alternate solution. The same geometry format as Figure 4 for the main solution.

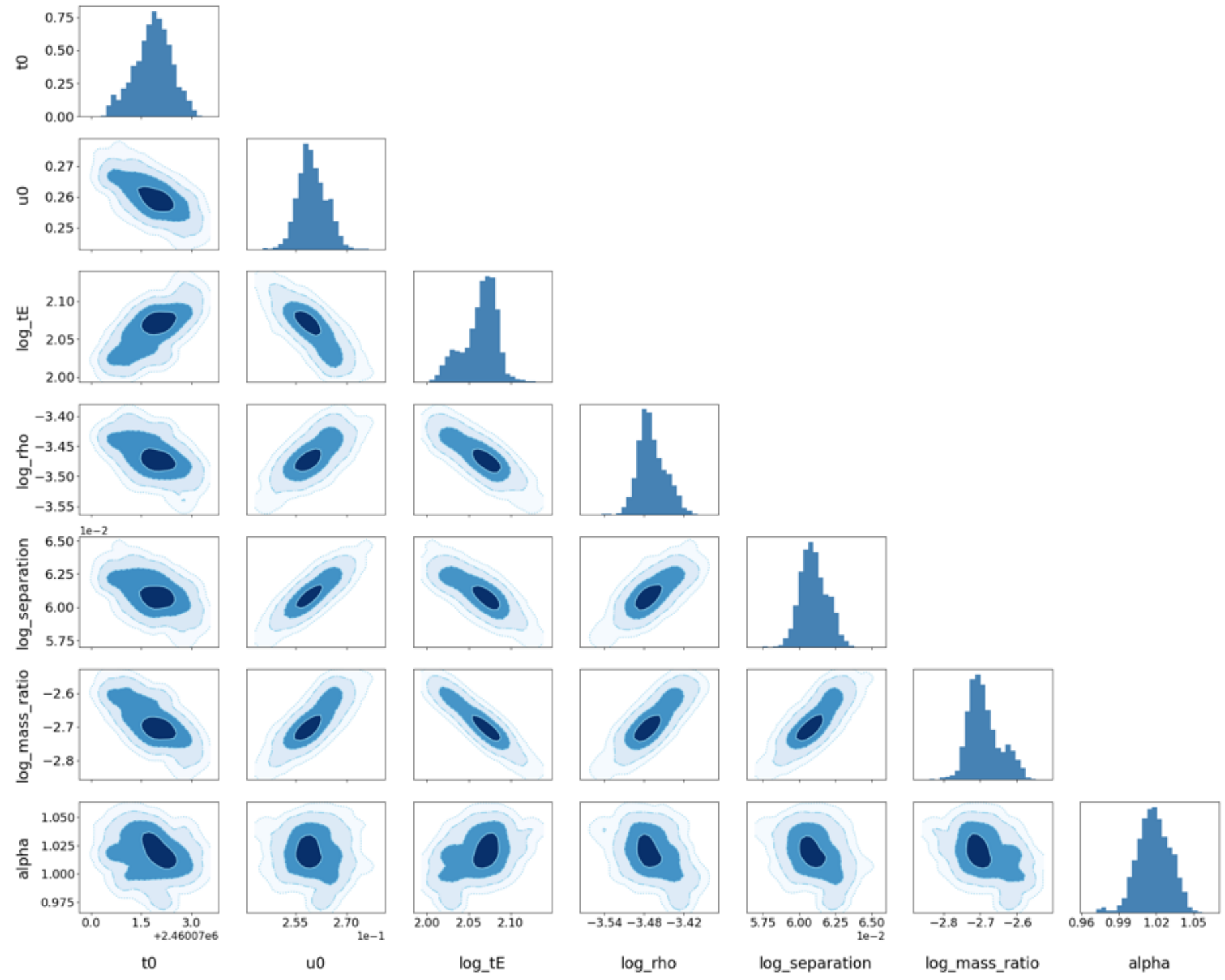

**Figure B3.** Corner plot of posterior distributions for the alternate model parameters of the `pyLIMA` fit depicted in Figure B1.

## Appendix C
## Limited Parallax Solution

The limited parallax solution reaches the imposed parameter bounds and does not match the photometry as well as the static FSBL model (see Section 3.4.2). We include the light curve with this model overlayed, (Figure C1) , the geometry predicted fion the lens refrence frame (Figure C2), and the corner plot showing the posteriors from the unfavored fit (Figure C3).

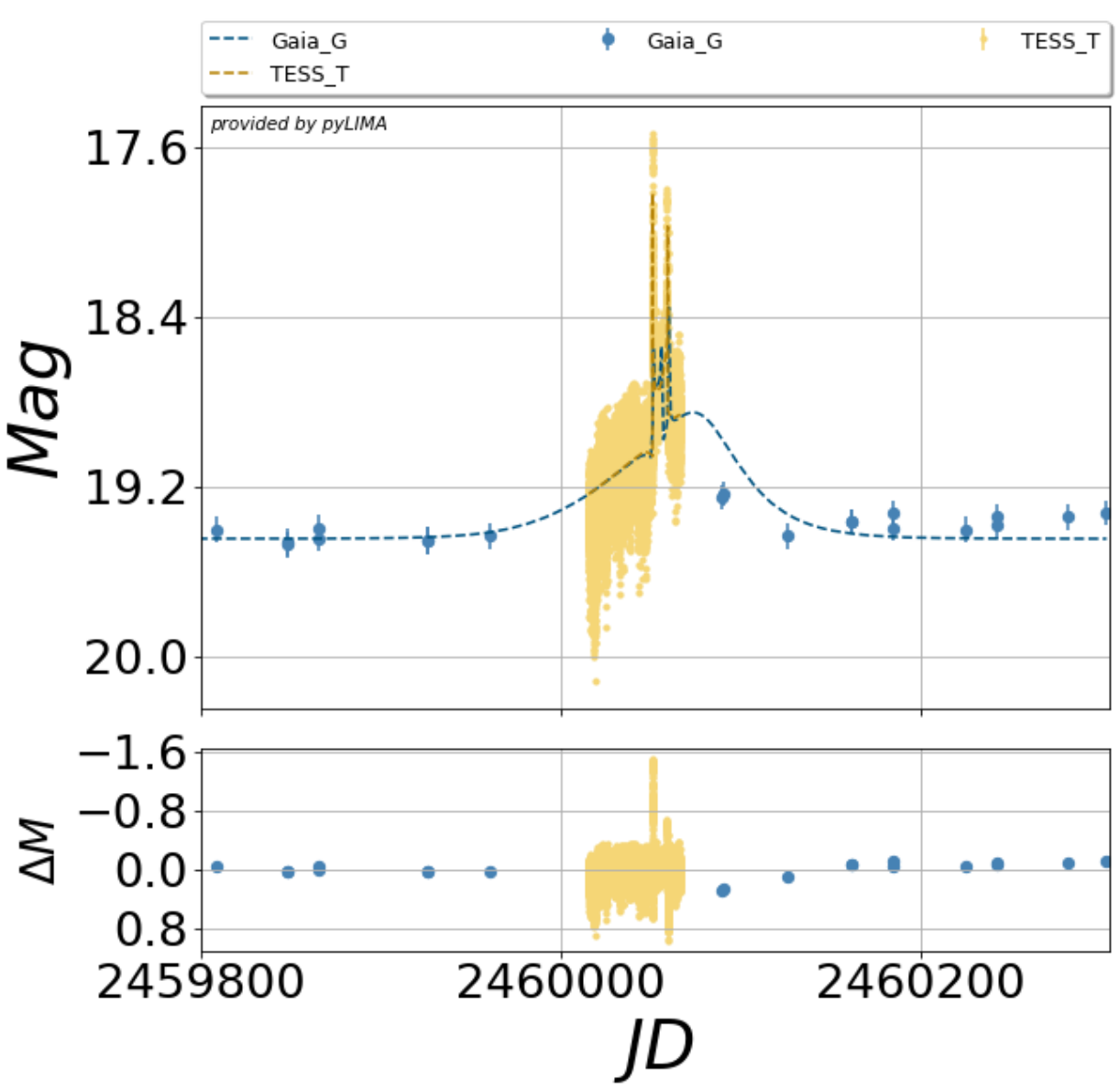

**Figure C1.** The photometry with the model overlaid for the limited parallax solution. While this solution is not as strong a fit as the static solution, and both parallax measures approach the upper limits of their bounds, a greater parallax measurement would disagree with the observed photometry. We dismiss this solution, and deduce that there is not enough information from this event to measure the parallax. The format is identical to Figure 2.

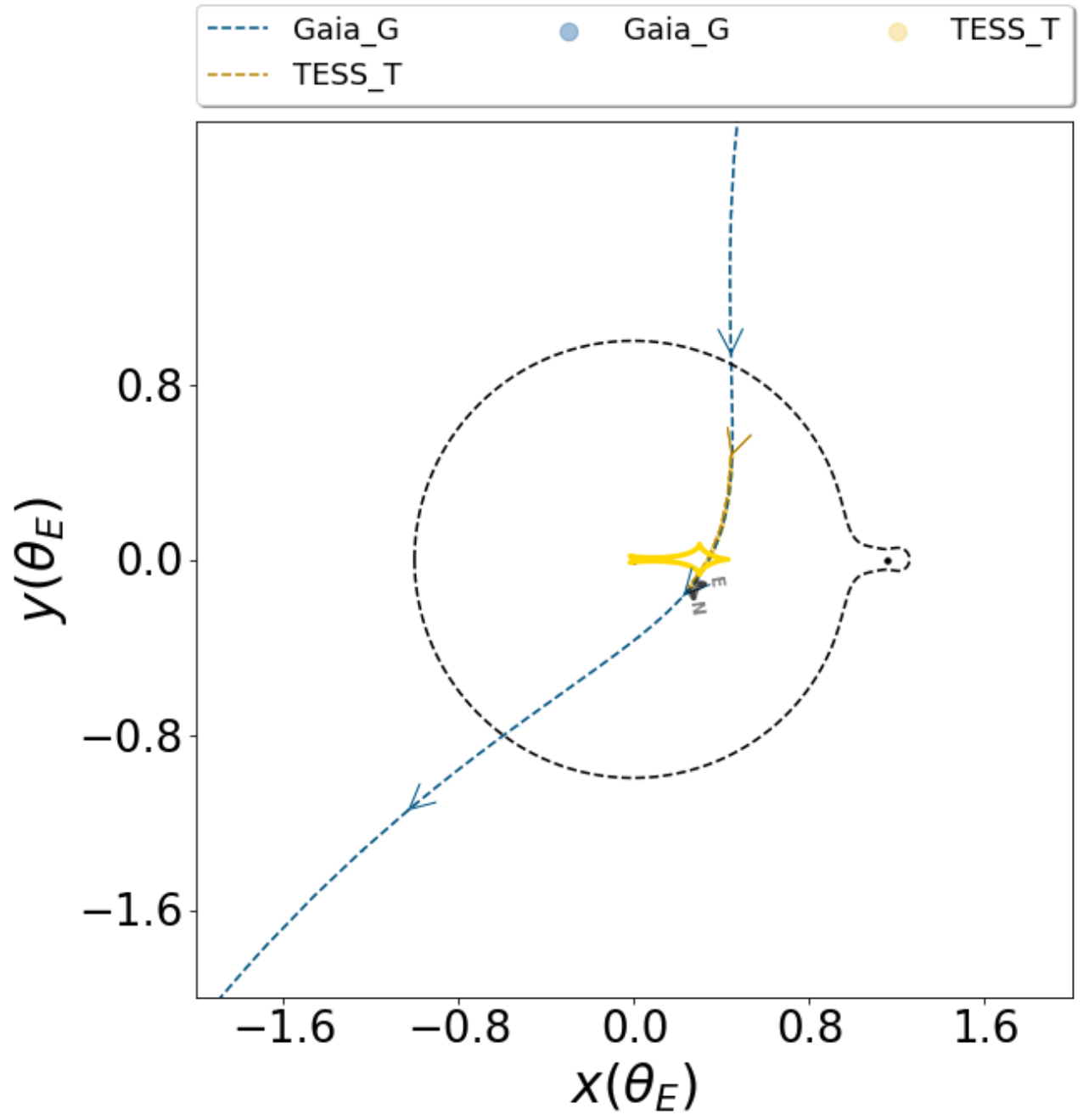

**Figure C2.** Geometry of the system from pyLIMA for the limited parallax solution. The same geometry format as Figure 4 for the main solution.

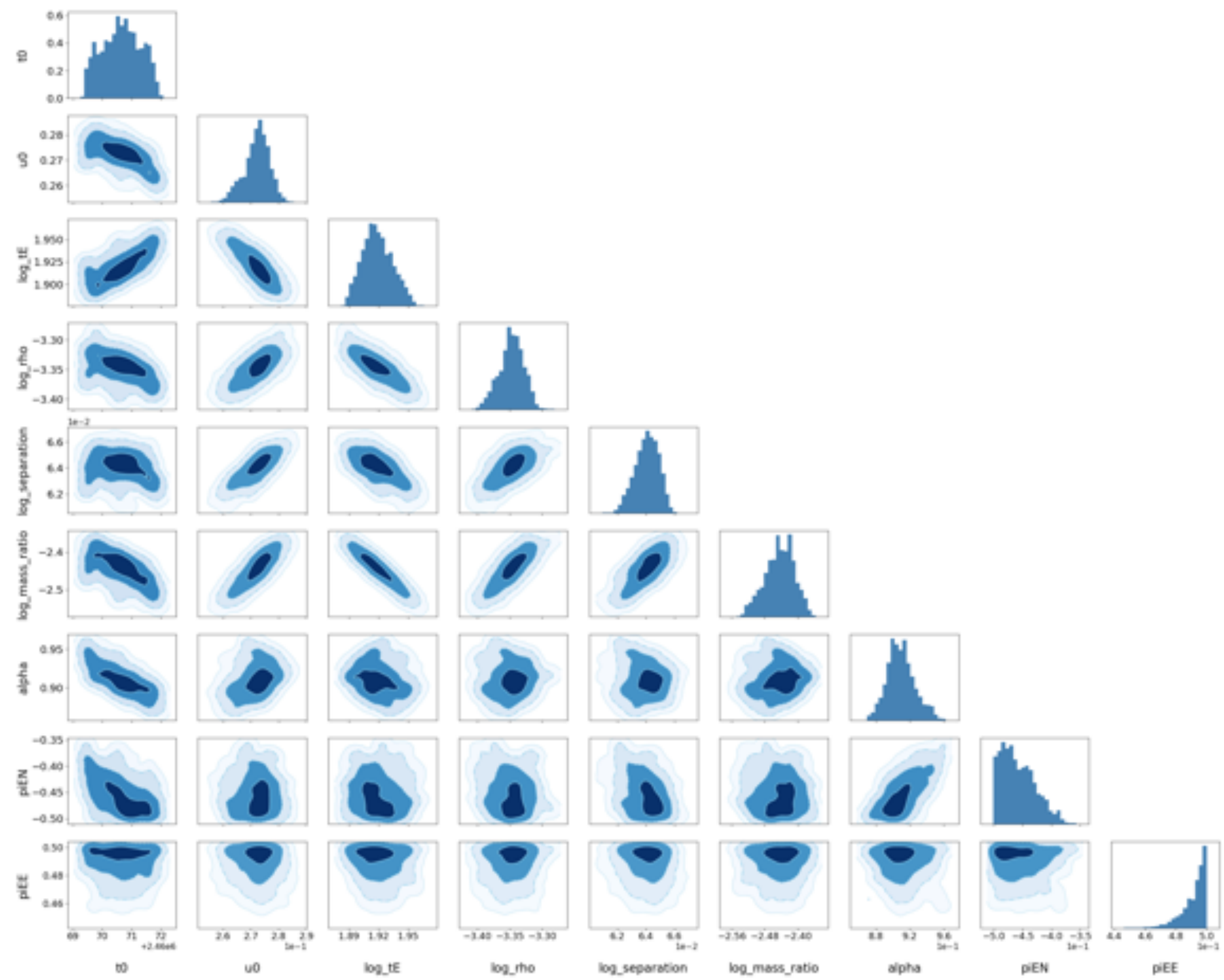

**Figure C3.** Corner plot of posterior distributions for the limited parallax model parameters of the `pyLIMA` fit depicted in Figure C1. The same parameters as Figure 3, with the addition of $\pi_{\rm EN}$ and $\pi_{\rm EE}$.

## ORCID iDs

Mallory Harris https://orcid.org/0000-0002-3721-1683
Diana Dragomir https://orcid.org/0000-0003-2313-467X
Etienne Bachelet https://orcid.org/0000-0002-6578-5078
Michael Fausnaugh https://orcid.org/0000-0002-9113-7162
Samson Johnson https://orcid.org/0000-0001-9397-4768